\documentclass[prd,amsmath,superscriptaddress,twocolumn]{revtex4}

\usepackage{graphicx}
\usepackage{amsmath}
\usepackage{amssymb}
\usepackage{graphics}
\usepackage{hyperref}
\usepackage{mathrsfs}
\usepackage{bm}
\usepackage[usenames,dvipsnames]{color}

\newcommand{\Bdtd}{B_{\delta\theta\delta}}
\newcommand{\Bdtt}{B_{\delta\theta\theta}}
\newcommand{\Btdd}{B_{\theta\delta\delta}}
\newcommand{\Btdt}{B_{\theta\delta\theta}}
\newcommand{\Bttd}{B_{\theta\theta\delta}}
\newcommand{\Bttt}{B_{\theta\theta\theta}}

\newcommand{\dtt}{{\delta\theta\theta}}

\newcommand{\ttd}{{\theta\theta\delta}}
\newcommand{\ttt}{{\theta\theta\theta}}
\newcommand{\Pdd}{P_{\delta\delta}}
\newcommand{\Pdt}{P_{\delta\theta}}

\newcommand{\Ptt}{P_{\theta\theta}}

\newcommand{\Btns}{B_\mathrm{TNS}}
\newcommand{\Pbis}{P_\mathrm{B}}
\newcommand{\Pbisj}[1]{P_{\mathrm{B,}#1}}
\newcommand{\Ptri}{P_\mathrm{T}}
\newcommand{\PLl}{P_{{\mathrm L,}\ell}}
\newcommand{\plustominus}{[\vec p_+ \leftrightarrow \vec p_-]}
\newcommand{\Ptrij}[1]{P_{\mathrm{T,}#1}}
\newcommand{\Qlabc}{Q_{abc}^{(\ell)}}
\newcommand{\QQ}[2]{Q^{(#1)}_{#2}}
\newcommand{\Rlabc}{R_{abc}^{(\ell)}}
\newcommand{\ttns}{\tau_\mathrm{TNS}}
\newcommand{\TNS}{\cite{Taruya_etal_2010}}

\newcommand{\BetaP}{{\mathscr B}_\nu}
\newcommand{\Class}{{\tt{CLASS}}}
\newcommand{\Copter}{{\tt{Copter}}}
\newcommand{\deltacb}{\delta_\mathrm{CB}}
\newcommand{\deltam}{\delta_\mathrm{m}}
\newcommand{\deltanu}{\delta_\nu}
\newcommand{\Dirac}{\delta_\mathrm{D}}
\newcommand{\fcb}{f_\mathrm{CB}}
\newcommand{\Ffog}{F_\mathrm{fog}}
\newcommand{\fnu}{f_\nu}
\newcommand{\Hc}{{\mathcal H}}
\newcommand{\lincb}{{\mathrm{lin,CB}}}
\newcommand{\linnu}{{{\mathrm{lin,}}\nu}}
\newcommand{\muv}[1]{\mu_{\vec #1}}
\newcommand{\ns}{n_\mathrm{s}}
\newcommand{\omegab}{\omega_\mathrm{b}}
\newcommand{\Omegab}{\Omega_\mathrm{b}}
\newcommand{\Omegam}{\Omega_\mathrm{m}}
\newcommand{\omegam}{\omega_\mathrm{m}}
\newcommand{\omeganu}{\omega_\nu}
\newcommand{\Omeganu}{\Omega_\nu}
\newcommand{\OmegaTRG}{{\bf \Omega}}
\newcommand{\Pleg}{{\mathscr P}}
\newcommand{\PCB}{P_\mathrm{CB}}
\newcommand{\Plin}{P_\mathrm{lin}}
\newcommand{\Pmat}{P_\mathrm{m}}
\newcommand{\Pnu}{P_\nu}
\newcommand{\rhobarcb}{{\bar \rho}_\mathrm{CB}}
\newcommand{\rhobarm}{{\bar \rho}_\mathrm{m}}
\newcommand{\rhobarnu}{{\bar \rho}_\nu}
\newcommand{\rhom}{\rho_\mathrm{m}}
\newcommand{\zin}{z_\mathrm{in}}

\newcommand{\MoooNo}{$\Lambda$CDM}
\newcommand{\MoooNi}{$\nu\Lambda$CDM}
\newcommand{\MooiNo}{EDE}
\newcommand{\MoiiNi}{$\nu$wCDM}

\begin{document}

\title{Redshift-space distortions in massive neutrino and evolving dark energy cosmologies} 

\author{Amol Upadhye}
\affiliation{Department of Physics, 
  University of Wisconsin--Madison, 
  1150 University Avenue, Madison, WI 53706}
\author{Juliana Kwan}
\affiliation{Department of Physics and Astronomy,
  University of Pennsylvania, Philadelphia, PA 19104}
\author{Adrian Pope}
\affiliation{High Energy Physics Division, Argonne National
  Laboratory, 9700 S. Cass Ave., Lemont, IL 60439} 
\author{Katrin Heitmann}
\author{Salman Habib}
\affiliation{High Energy Physics Division, Argonne National
  Laboratory, 9700 S. Cass Ave., Lemont, IL 60439}  
\affiliation{Kavli Institute for Cosmological Physics, The University
  of Chicago, Chicago, IL 60637}
\affiliation{Mathematics and Computer Science Division, Argonne
  National Laboratory, 9700 S. Cass Ave., Lemont IL 60439}
\author{Hal Finkel}
\affiliation{ALCF, Argonne National Laboratory, 9700 S. Cass Ave.,
  Lemont, IL 60439}%
\author{Nicholas Frontiere}
\affiliation{High Energy Physics Division, Argonne National
  Laboratory, 9700 S. Cass Ave., Lemont, IL 60439} 
\affiliation{Department of Physics, The University of Chicago,
  Chicago, IL 60637} 

\date{\today}

\begin{abstract}
Large-scale structure surveys in the coming years will measure the redshift-space power spectrum to unprecedented accuracy, allowing for powerful new tests of the $\Lambda$CDM picture as well as measurements of particle physics parameters such as the neutrino masses.  We extend the Time-RG perturbative framework to redshift space, computing the power spectrum $P_s(k,\mu)$ in massive neutrino cosmologies with time-dependent dark energy equations of state $w(z)$.  Time-RG is uniquely capable of incorporating scale-dependent growth into the $P_s(k,\mu)$ computation, which is important for massive neutrinos as well as modified gravity models.  Although changes to $w(z)$ and the neutrino mass fraction both affect the late-time scale-dependence of the non-linear power spectrum, we find that the two effects depend differently on the line-of-sight angle $\mu$.  Finally, we use the HACC N-body code to quantify errors in the perturbative calculations.  For a $\Lambda$CDM model at redshift $z=1$, our procedure predicts the monopole~(quadrupole) to $1\%$ accuracy up to a wave number $0.19h/$Mpc~($0.28h/$Mpc), compared to $0.08h/$Mpc~($0.07h/$Mpc) for the Kaiser approximation and $0.19h/$Mpc~($0.16h/$Mpc) for the current state-of-the-art perturbation scheme.  Our calculation agrees with the simulated redshift-space power spectrum even for neutrino masses above the current bound, and for rapidly-evolving dark energy equations of state, $|dw/dz| \sim 1$.  Along with this article, we make our redshift-space Time-RG implementation publicly available as the code {\tt{redTime}}.
\end{abstract}

\maketitle

\section{Introduction}
\label{sec:introduction}

Two major challenges for cosmology over the next decade are measuring the neutrino masses and constraining the evolution of the dark energy density.  The sum of the neutrino masses, a fundamental Standard Model parameter, is bounded from above by cosmological probes~\cite{Ade_2015xiii,Anderson_2014}: $\sum m_\nu < 0.23$~eV.  Improved measurements of large-scale structure and the cosmic microwave background over the next several years will replace this bound with a measurement, possibly allowing us to distinguish between normal and inverted neutrino mass hierarchies~\cite{Font-Ribera_2014,Schlegel_2009,Levi_2013,Abbott_2005,Abell_2009,Refregier_2010,Basse_2014}.

Meanwhile, searches for dark energy evolution are entering a decisive era.  The cosmological constant $\Lambda$, the simplest model of dark energy, is completely consistent with current data~\cite{Vikhlinin_2009,Conley_2011,Suzuki_etal_2012,Kilbinger_2013,Hinshaw_2013,Hou_2014,Calabrese_2013,Ade_etal_2013xvi, Anderson_2014,Ade_2015xiii}.  However, a cosmological energy density which is $120$ orders of magnitude below fundamental scales, yet coincidentally nearly equal to the dark matter density today, requires a great deal of fine-tuning~\cite{Weinberg_1989,Bousso_2000,Nobbenhuis_2006,Caldwell_2009,Martin_2012}.  If we consider models in which the dark energy density $\rho_\mathrm{DE}$ and its equation of state $w_\mathrm{DE}(z) = P_\mathrm{DE} / \rho_\mathrm{DE}$ may vary with redshift $z$, the uncertainty in $dw_\mathrm{DE}/dz$ is of order unity today~\cite{Ade_2015xiv}.  Current constraints allow ``early dark energy'' models in which $\rho_\mathrm{DE}(z)$ rises rapidly with $z$, substantially alleviating the tuning and coincidence problems associated with the cosmological constant~\cite{Binetruy_1999,Zlatev_1999,Steinhardt_1999,Armendariz-Picon_2000,Barriero_2000,Upadhye_Ishak_Steinhardt_2005,Bueno_Sanchez_2006,Alam_2010,Alam_2011}.  Over the next ten to fifteen years, constraints on $dw_\mathrm{DE} / dz$ will improve substantially, allowing models with $|dw_\mathrm{DE} / dz| \sim 1$ to be distinguished decisively from slowly-evolving equations of state~\cite{Font-Ribera_2014}.  Thus cosmology is poised to answer two fundamental questions about the nature of the universe.

Such powerful cosmological constraints will be made possible by combining probes of the expansion rate $H(z)$, including Type Ia supernova~\cite{Betoule_2013} and baryon acoustic oscillation (BAO)~\cite{Anderson_2014} surveys, with measurements of the growth factor $D(z)$, including redshift-space distortions (RSD)~\cite{Beutler_2014,Samushia_2014} and weak lensing~\cite{Heymans_2013}.  Here we are particularly interested in the RSD, which probe the logarithmic growth rate $f(z) = -d\log(D) / d\log(1+z)$.  These are measured at quasi-linear scales $10$~Mpc -- $100$~Mpc, and they are minimally affected by astrophysical systematics such as baryonic feedback.  As a result, perturbative treatments of the redshift-space power spectrum are feasible~\cite{Kaiser_1987,Scoccimarro_2004,Taruya_etal_2010,Kwan_2012}.

Time-RG perturbation theory, which directly integrates a system of non-linear equations for the matter power spectrum $P(k)$ (the Fourier transform of the two-point correlation function), was designed for models with scale-dependent growth factors, such as massive neutrino, modified gravity, and clustering dark energy models~\cite{Pietroni_2008,Lesgourgues_etal_2009,Anselmi_2011}.  It is implemented in publicly-available codes including~\Copter~\cite{Carlson_White_Padmanabhan_2009} and~\Class~\cite{Audren_2011}.  Its approach to computing the power spectrum is to truncate the infinite tower of evolution equations for $N$-point correlation functions.  Since the continuity and Euler equations of classical fluid dynamics relate the time derivative of the $N$-point function to the ($N+1$)-point function, the power spectrum $P(\vec k)$ depends upon the bispectrum $B(\vec k_1, \vec k_2, \vec k_3)$,  the bispectrum upon the trispectrum, and so on.  Time-RG truncates this hierarchy by neglecting the connected part of the trispectrum, allowing the bispectrum to describe the non-linear evolution of the power spectrum.

Time-RG results were compared with those from N-body simulations in~\cite{Carlson_White_Padmanabhan_2009,Upadhye_2014}.  In particular, Ref.~\cite{Upadhye_2014} showed that Time-RG accurately predicted the power spectrum even for early dark energy models which cause Standard Perturbation Theory to fail, and for massive neutrino models to which other perturbative treatments are inapplicable.  Furthermore, Time-RG was used to test the approximation of Refs.~\cite{Saito_2008,Agarwal_2011}, in which neutrino clustering is neglected as a source for non-linear dark matter growth.

In this article, we extend Time-RG to a prediction of the redshift-space power spectrum using the approach of Ref.~\TNS, which describes higher-order corrections to the redshift-space power spectrum in terms of integrals $\Pbis(k,\mu)$ and $\Ptri(k,\mu)$ over the bispectrum and trispectrum, respectively.  In particular, we show that $\Pbis(k,\mu)$ can be expressed as a linear combination of terms depending on $k$ alone, whose time-evolution can be computed in the Time-RG framework.  Our treatment automatically includes corrections to $\Pbis$ due to non-linear evolution, and we show that these corrections result in a smearing of baryon oscillations in the $\Pbis$ terms as well as a transfer in power from larger to smaller scales.  Extending this calculation to massive neutrino models, we show that $\sum m_\nu$ and the equation of state parameters affect the redshift-space power spectrum differently, due to the scale-dependent growth sourced by massive neutrinos.

Next, we compare our calculations of the redshift-space power spectrum to the results of the HACC high-precision N-body simulations~\cite{Habib_2014} used in Ref.~\cite{Upadhye_2014}.  For models without massive neutrinos, we show that the Time-RG predictions remain accurate down to smaller scales than those of Ref.~\TNS, largely due to the better small-scale behavior of Time-RG relative to closure perturbation theory.  Finally, we demonstrate that our approach accurately computes the redshift-space power spectrum for the full range of $\sum m_\nu$ allowed by current data~\cite{Ade_2015xiii,Wyman_2014}, and for models with cosmological constants as well as rapidly-varying dark energy.

This article is organized as follows.  Section~\ref{sec:background} covers Time-RG perturbation theory, the linear neutrino approximation used here, and the basics of perturbative RSD calculations.  Our new results are derived in Sec.~\ref{sec:rsd_in_trg}, and after comparing them with the calculations of~\TNS~we contrast the effects of $\sum m_\nu$ and the equation of state parameters on the redshift-space power spectrum.  N-body simulations are used to test our results in Sec.~\ref{sec:comparison_with_simulations}. Section~\ref{sec:conclusions} concludes that our redshift-space Time-RG calculation agrees closely with N-body simulations in massive neutrino and evolving dark energy cosmologies, and discusses possible applications.

\section{Background}
\label{sec:background}

\subsection{Time-RG Perturbation Theory} 
\label{subsec:time-rg_perturbation_theory}

Consider a spatially-flat universe containing several non-relativistic fluids, each with density $\rho_I(\tau,\vec x)$ and velocity field $\vec v_I(\tau,\vec x)$, where $\tau$ and $\vec x$ are the conformal time and comoving position, respectively.~\footnote{We do not consider models with spatial curvature in this article, and we note that the fluid approximation itself breaks down due to multi-streaming at small scales, limiting the reach of fluid-based perturbation theories.}  Assume that the fluids interact only gravitationally.  In terms of the density contrasts $\delta_I(\tau,\vec x) = \rho_I(\tau,\vec x)/\bar\rho_I(\tau) - 1$, where the $\bar\rho_I(\tau)$ are the mean densities, we can write down the continuity and Euler equations for each fluid as well as a Poisson equation coupling them:
\begin{eqnarray}
\frac{\partial \delta_I}{\partial\tau}
+
\vec\nabla \cdot \vec v_I
+
\vec\nabla\cdot (\vec v_I \delta_I)
&=&
0
\label{e:continuity}
\\
\frac{\partial \vec v_I}{\partial\tau}
+
\Hc \vec v_I
+
(\vec v_I \cdot \vec\nabla)\vec v_I
+
\vec\nabla \Phi
&=&
0
\label{e:Euler}
\\
\nabla^2 \Phi 
- 
\frac{3}{2} \Omegam(\tau) \Hc^2 \sum_J f_J \delta_J
&=&
0
\label{e:Poisson}
\end{eqnarray}
where the total cold matter density $\rhom = \sum_J f_J \rho_J$.  Here $\Phi$ is the gravitational potential and $\Hc = a^{-1} da/d\tau$ the conformal Hubble rate.  In the regime of validity of cosmological perturbation theory, the velocity fields are well-approximated as irrotational, $\vec \nabla \times \vec v_i = 0$.  Then each velocity can be described completely by its divergence $\theta_I = \Hc^{-1} {\vec \nabla} \cdot \vec v_i$ using $\nabla^2 \vec v_I = \Hc \vec \nabla \theta_I$.  Henceforth we describe each fluid in terms of these scalar perturbations $\delta_I$ and $\theta_I$.

We may put the equations of motion into a more compact notation.  Let $\eta = \log[(1+\zin)/(1+z)]$, where $\zin \gg 1$ is an initial redshift at which non-linear terms in the equations of motion are negligible.  Working in redshift space, define $\varphi_{I0} = \exp(-\eta)\delta_I$ and $\varphi_{I1} = -\exp(-\eta)\theta_I$.  With primes ($'$) denoting derivatives with respect to $\eta$, and summation over repeated lower-case indices, we have
\begin{eqnarray}
\varphi_{Ia}'
+
\OmegaTRG_{Iab} \varphi_{Ib}
&=&
e^\eta
\int\frac{d^3q}{(2\pi)^3} \frac{d^3p}{(2\pi)^3}
\gamma_{abc}(\vec k, -\vec p, -\vec q)\qquad
\label{e:eom_PT}
\\
&~&
\quad
\times
\varphi_{Ib}(\vec p) \varphi_{Ic}(\vec q)
(2\pi^3)\Dirac^3(\vec k + \vec p + \vec q)
\nonumber\\
\OmegaTRG_{I00} = -\OmegaTRG_{I01}
&=&
1
\\
\OmegaTRG_{I10}
&=&
-\frac{3}{2}
\Omegam(\eta)
\left(f_I + \sum_{J \neq I} f_J \frac{\delta_J}{\delta_I} \right)
\label{e:eom_Omega10}
\\
\OmegaTRG_{I11}
&=&
2 + \Hc'/\Hc
\\
\gamma_{001}(\vec k,\vec q, \vec p) 
&=& 
\gamma_{010}(\vec k, \vec p, \vec q) 
= 
(\vec p + \vec q)\cdot \vec q / (2p^2)
\\
\gamma_{111}(\vec k, \vec q, \vec p) 
&=& 
(\vec p + \vec q)^2 \vec p \cdot \vec q / (2p^2q^2)
\end{eqnarray}
with all other $\gamma_{abc}$ zero.  In the first line, the $\vec k$-dependence of $\varphi$ on the left-hand-side  has been suppressed, and $\Dirac$ is the Dirac delta function.  

Next consider a universe with a single fluid.  Using Eq.~(\ref{e:eom_PT}), Time-RG Perturbation Theory~\cite{Pietroni_2008} writes down the equations of motion for the power spectrum $P_{ab}(\vec k_1) (2\pi)^3 \Dirac^3(\vec k_1 + \vec k_2) = \left< \varphi_a(\vec k_1) \varphi_b(\vec k_2) \right>$ and bispectrum $B_{abc}(\vec k_1,\vec k_2, \vec k_3) (2\pi^3)\Dirac^3(\vec k_1 + \vec k_2 + \vec k_3) = \left< \varphi_a(\vec k_1) \varphi_b(\vec k_2) \varphi_c(\vec k_3) \right>$:
\begin{eqnarray}
P_{ab}'
&=&
-\Omega_{ac} P_{cb}(k) - \Omega_{bc}P_{ac}(k)
\nonumber\\&&
+ e^\eta \int\frac{d^3q}{(2\pi)^3} 
[\gamma_{acd}(k,q,p) B_{bcd}(k,q,p)
\nonumber\\&&
\qquad\qquad\quad + \gamma_{bcd}(k,q,p) B_{acd}(k,q,p)]
\label{e:eom_full_P}
\\
B_{abc}'
&=&
-\Omega_{ad}(k)B_{dbc} - \Omega_{bd}(q)B_{adc} - \Omega_{cd}(p) B_{abd}
\nonumber\\&&
+ 2e^\eta [ \gamma_{ade}(k,q,p) P_{db}(q) P_{ec}(p)
\nonumber\\&&
\qquad  + \gamma_{bde}(q,p,k) P_{dc}(p) P_{ea}(k)
\nonumber\\&&
\qquad  + \gamma_{cde}(p,k,q) P_{da}(k) P_{eb}(q)]
\label{e:eom_full_B}
\end{eqnarray}
where the bispectrum in the second equation is understood to be a function of $k$, $q$, and $p$: $B_{abc} = B_{abc}(k,q,p)$.

Needing to keep track of the functional dependence of $B_{abc}$ on three different variables would make Time-RG perturbation theory numerically prohibitive.  However, it is possible to recast the equations of motion in such a way that the time-dependent quantities depend only on one variable, $k$:
\begin{eqnarray}
P_{ab}'
&=&
-\Omega_{ac} P_{bc} - \Omega_{bc} P_{ac}
\nonumber\\&&
+ e^\eta \frac{4\pi}{k}(I_{acd,bcd} + I_{bcd,acd})
\label{e:eom_P}
\\
\frac{I_{acd,bef}}{k}
&=&
\!\!
\int \!\!\frac{q^2 dq \sin\alpha d\alpha}{(2\pi)^3} 
\frac{\gamma_{acd}(k,q,p_-\!) B_{bef}(k,q,p_-\!)}{2}\qquad
\label{e:Iacdbef}
\\
I_{acd,bef}'
&=&
-\Omega_{bg} I_{acd,gef} - \Omega_{eg} I_{acd,bgf} 
\nonumber\\&&
- \Omega_{fg} I_{acd,beg}
+ 2e^\eta A_{acd,bef}
\label{e:eom_I}
\\
\frac{A_{acd,bef}}{k}
&=&
\int \frac{q^2 dq \sin\alpha d\alpha}{(2\pi)^3} 
\frac{\gamma_{acd}(k,q,p_-)}{2}
\nonumber\\&&
\times \Big[
  \gamma_{bgh}(k,q,p_-) P_{ge}(q) P_{hf}(p_-)
\nonumber\\&&
  \quad+ \gamma_{egh}(q,p_-,k) P_{gf}(p_-) P_{hb}(k)
\nonumber\\&&
  \quad+ \gamma_{fgh}(p_-,k,q) P_{gb}(k) P_{he}(q)
\Big].
\label{e:Aacdbef}
\end{eqnarray}
Here, $\vec p_\pm = \vec k \pm \vec q$, and the $P_{ab}$, $I_{acd,bef}$, and $A_{acd,bef}$ depend on $k$ unless otherwise specified.  Since most of the $\gamma_{abc}$ are zero, there are only $24$ nonzero components of $I_{acd,bef}$ at each $k$, only $14$ of which are independent.

\begin{figure*}[tp]
  \includegraphics[width=3.4in]{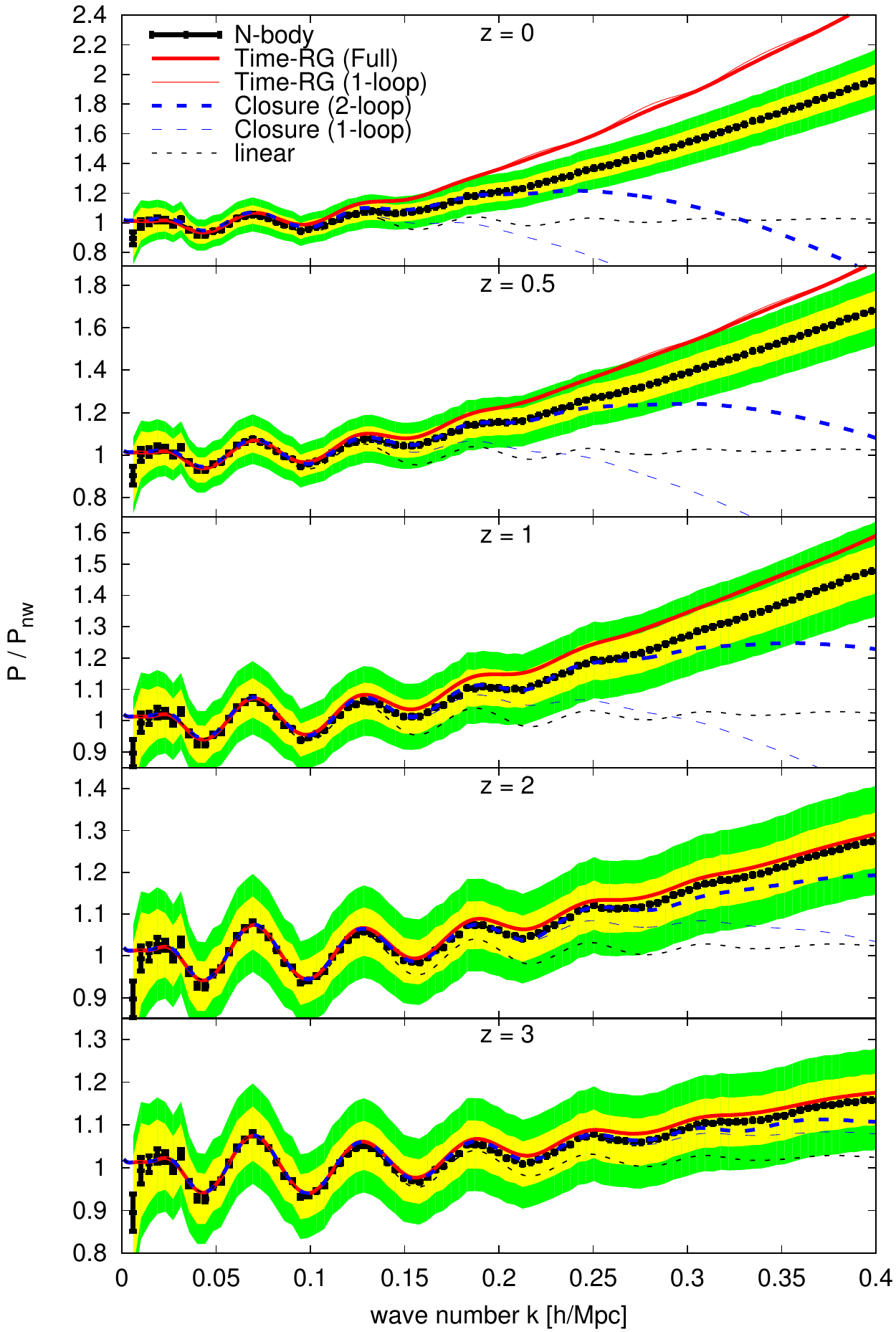}%
  \includegraphics[width=3.4in]{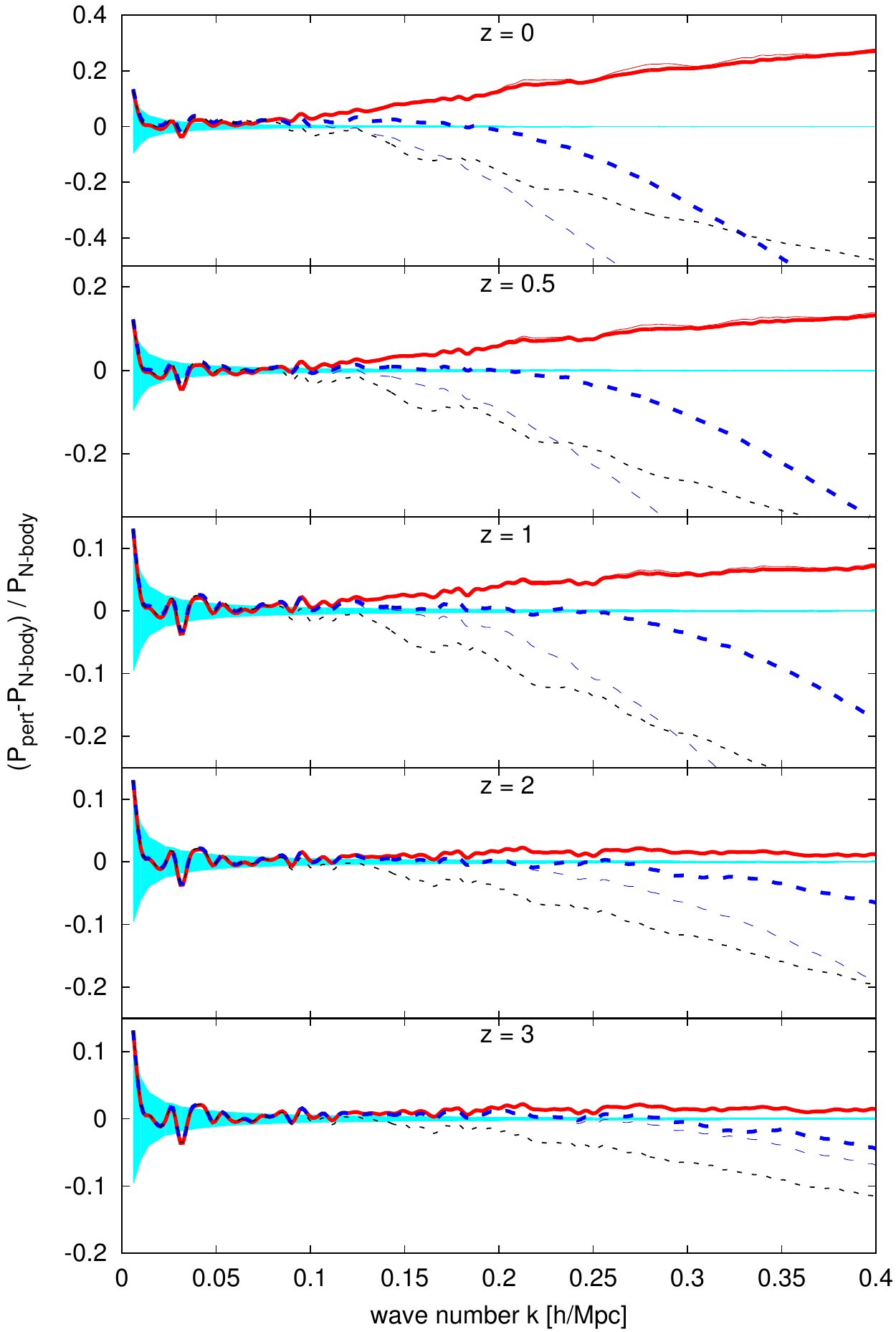}%
  \caption{
    Power spectrum $P_{\delta\delta}$ for the model~\MoooNo. 
    (Left)~Points with error bars show the N-body simulations of 
    Ref.~\cite{Upadhye_2014}, and
    the inner (outer) shaded regions are within $5\%$ ($10\%$) of the 
    simulation points.  Full and $1$-loop Time-RG are compared with the
    closure theory calculations of Ref.~\TNS.  Note that at the wave numbers 
    shown, full and $1$-loop Time-RG are indistinguishable except at $z=0$ and
    $k \gtrsim 0.25~h/$Mpc.
    (Right)~Fractional errors associated with the perturbation theories
    on the left.  The shaded region shows the $2\sigma$ N-body simulation
    error bars.
    \label{f:P00_M000n0}
  }
\end{figure*}

\begin{figure*}[tp]
  \includegraphics[width=3.4in]{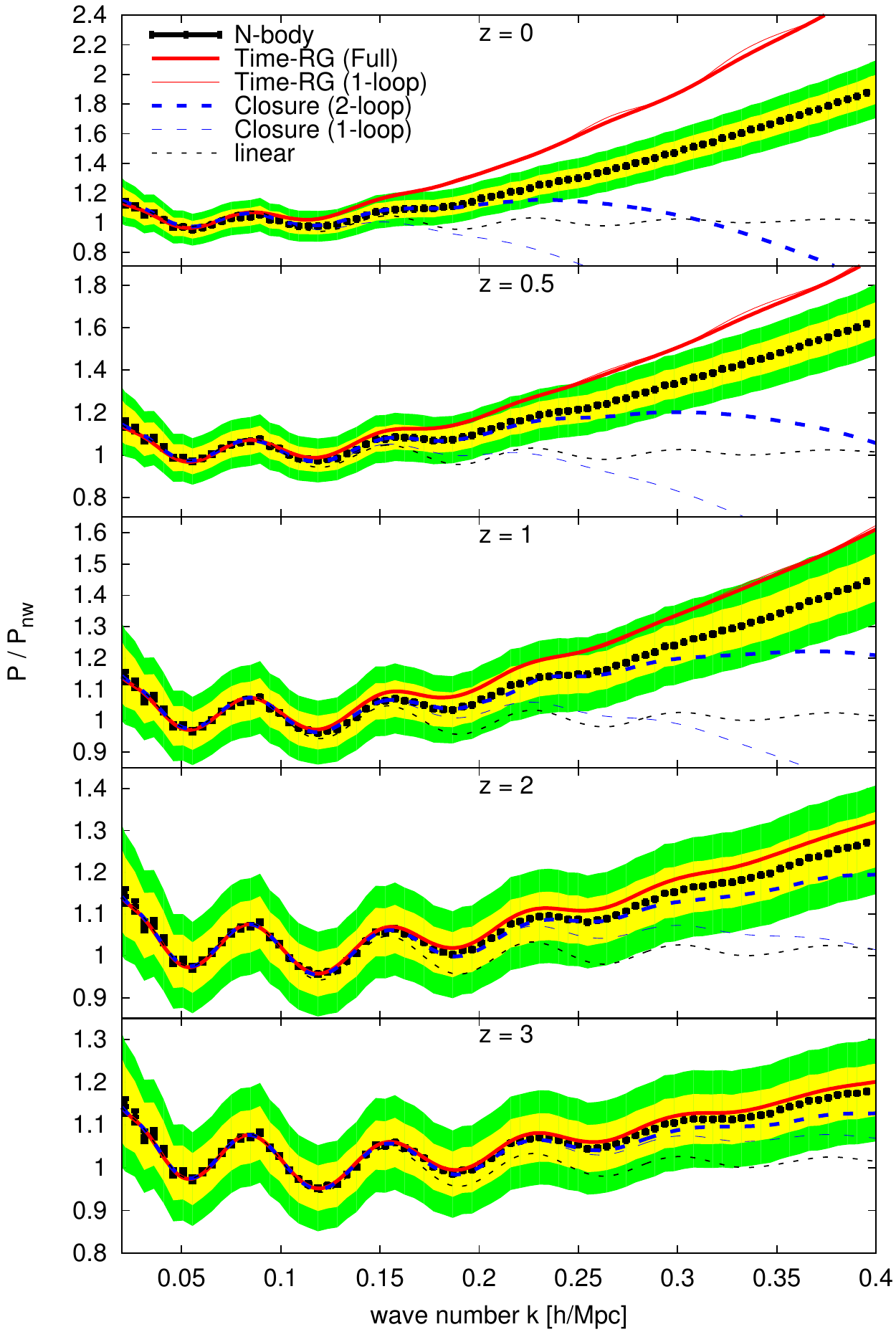}%
  \includegraphics[width=3.4in]{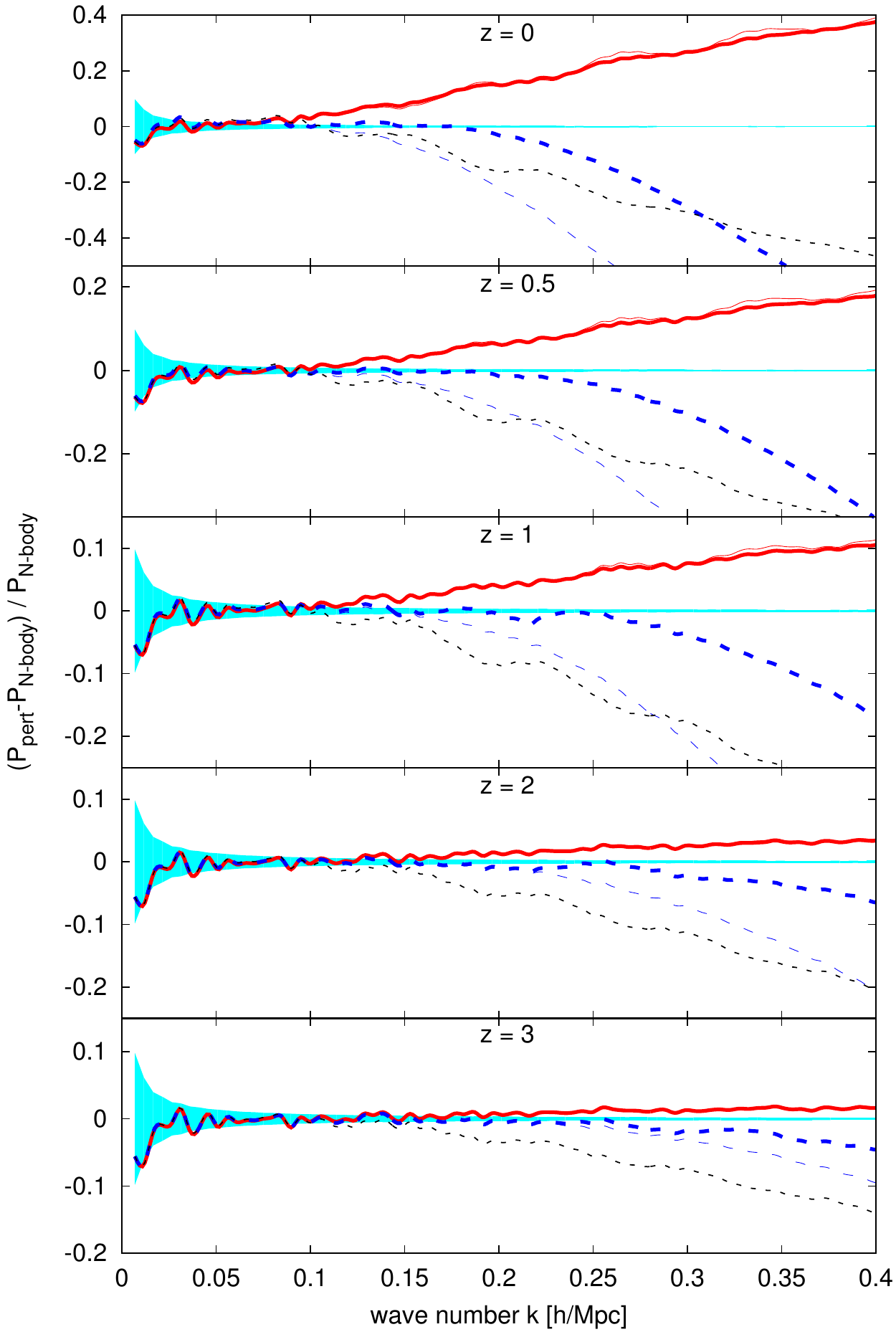}%
  \caption{
    Same as Figure~\ref{f:P00_M000n0}, but for the early dark energy
    model~\MooiNo~from Table~\ref{t:models}.
    \label{f:P00_M001n0}
  }
\end{figure*}

\begin{figure}[tp]
  \includegraphics[width=3.4in]{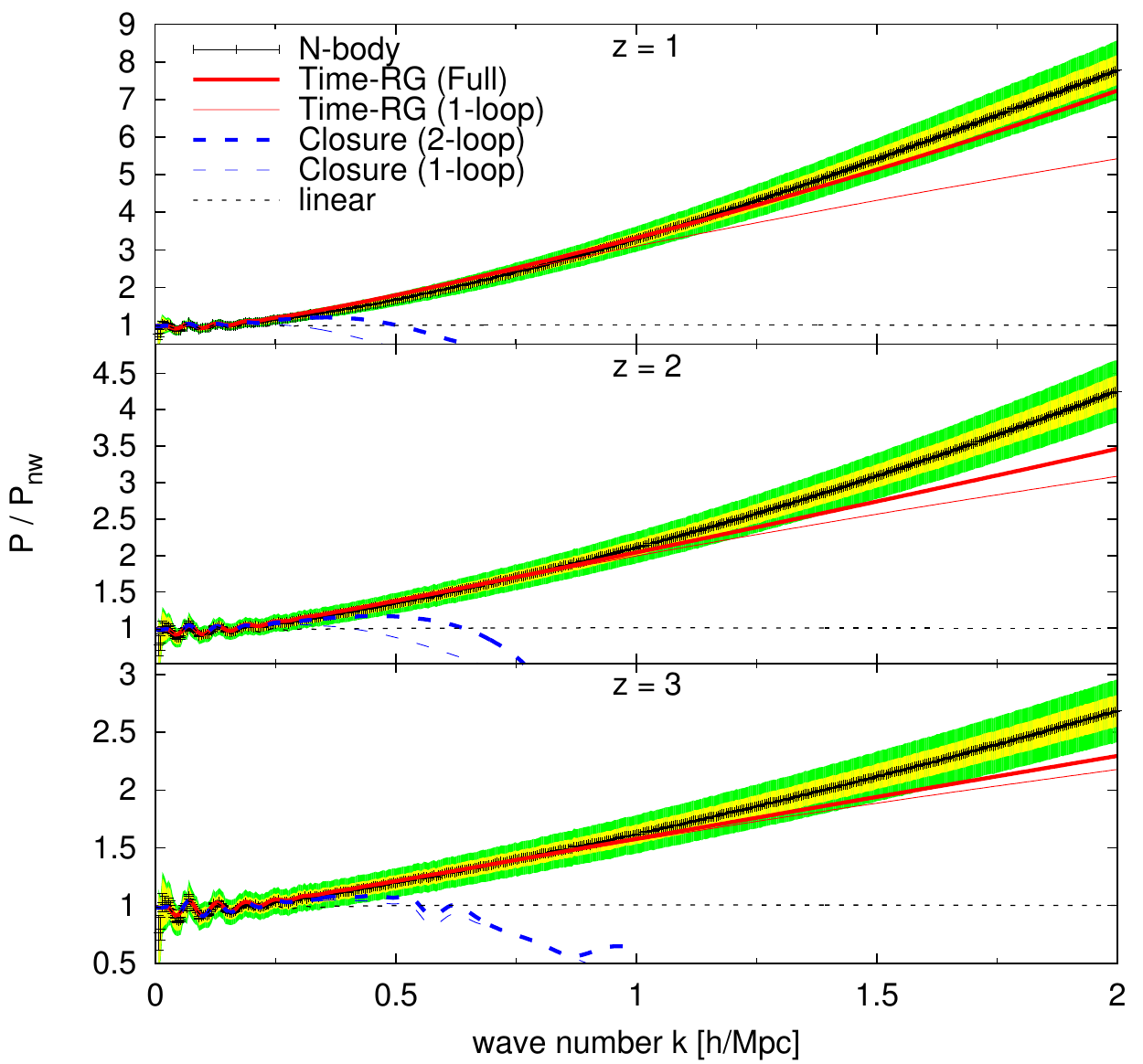}
  \caption{
    Same as Figure~\ref{f:P00_M000n0}, but using the high-resolution 
    TreePM simulation of~\cite{Upadhye_2014}, accurate to much smaller
    scales.
    \label{f:hires_M000n0}
  }
\end{figure}

\begin{figure}[tp]
  \includegraphics[width=3.4in]{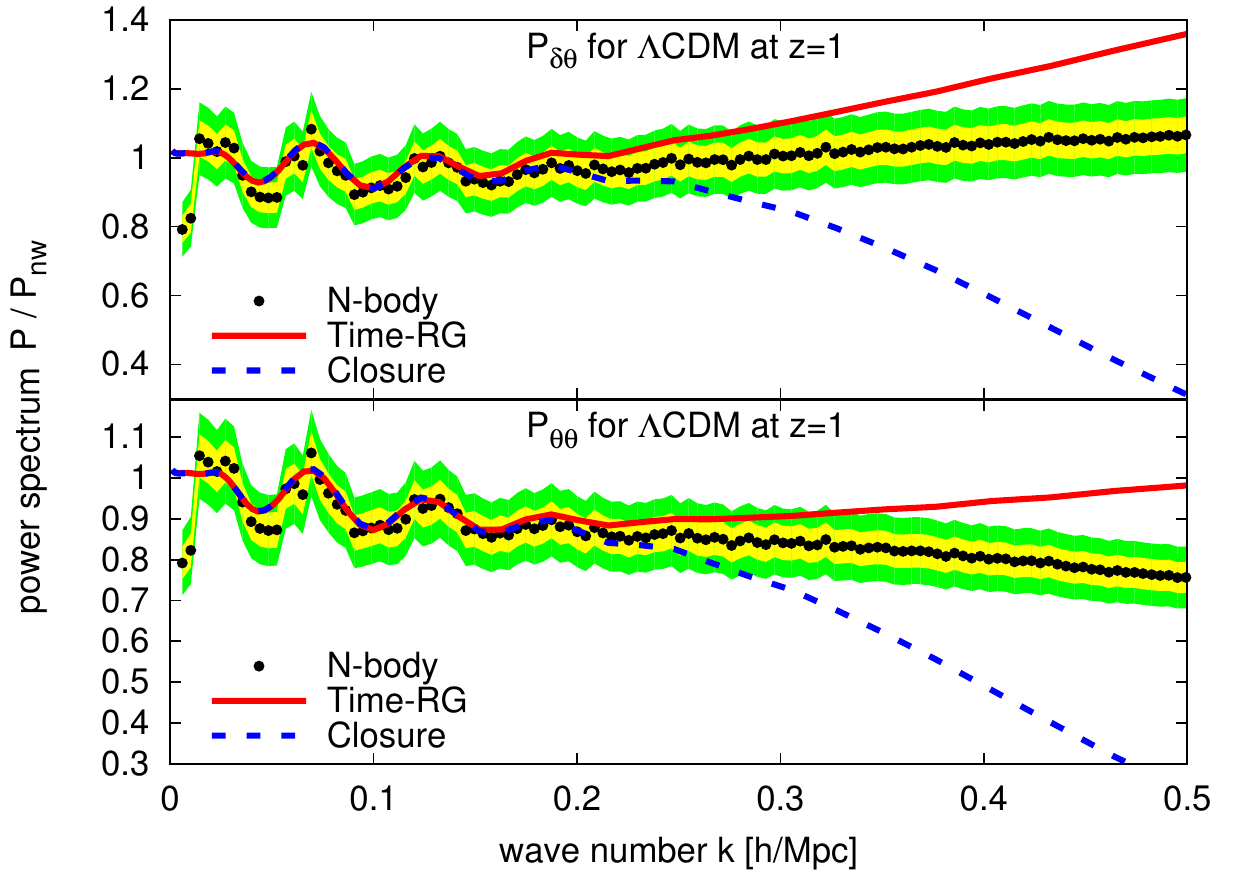}%
  \caption{
    Density-velocity cross spectrum $P_{\delta\theta}$ and velocity power spectrum 
    $P_{\theta\theta}$ at z=1 for~\MoooNo.  Inner and outer shaded regions
    correspond to $5\%$ and $10\%$ deviations, respectively, from the
    N-body points.
    \label{f:Pdv_Pvv_M000n0}
  }
\end{figure}

\begin{table}[tbp]
  \begin{footnotesize}
    \begin{tabular}{r|cccccccc}
      Model  & $h$      & $\omegam$ & $\omegab$ & $\omeganu$ & $\ns$    & $\sigma_8$ & $w_0$  & $w_a$\\
      \hline
      \MoooNo & $0.71$   & $0.1335$  & $0.02258$ & $0$        & $0.963$  & $0.8$      & $-1$   & $0$ \\
      \MoooNi & $0.71$   & $0.1335$  & $0.02258$ & $0.001$     & $0.963$  & $0.8$      & $-1$   & $0$ \\
     \MooiNo & $0.6167$ & $0.1472$  & $0.02261$ & $0$        & $0.9611$ & $0.8778$   & $-0.7$ & $0.6722$\\
      \MoiiNi & $0.7342$ & $0.1543$ & $0.02323$ & $0.003$ & $0.8797$ & $0.8056$ & $-1.21$ & $-1.11$\\
    \end{tabular}
  \end{footnotesize}
  \caption{
    List of cosmological models considered here.  For each component $J$, $\omega_J = \Omega_J(\tau_0) h^2$, where $\tau_0$ is the time today.
    \label{t:models}
  }
\end{table}

Before proceeding, we note that the treatment of non-linearities in Eq.~(\ref{e:eom_I}) neglects any scale-dependence in $\OmegaTRG$.  In particular, in a cosmological model containing multiple species which cluster differently, such as cold dark matter and massive neutrinos, $\OmegaTRG_{10}$ will depend on $k$ as given by Eq.~(\ref{e:eom_Omega10}).  Since this is a small correction to a non-linear correction, it is expected to be small~\cite{Blas_2014}, and it will be further suppressed when the density fraction of the second species is small.  We defer discussion of the scale dependence of Eq.~(\ref{e:eom_I}) and its redshift-space generalization to Appendix~\ref{sec:error_bounds_and_sigv_fits}, in which we show that the associated error is less than $1\%$ all the way to $z=0$ and $k=0.4~h/$Mpc even for neutrino masses $\sum m_\nu =0.94$~eV much larger than current bounds.

The Time-RG calculations of Eqs.~(\ref{e:eom_P}-\ref{e:Aacdbef}) may be sped up considerably by replacing the power spectra inside the integral of Eq.~(\ref{e:Aacdbef}) by the linear-theory power spectra.  Since these scale in a known way with the growth factor $D(z)$ and its logarithmic derivative $f(z) = -d\log D / d\log (1+z)$, the integral need not be repeated at each time step.  This is known as the ``$1$-loop'' approximation of Time-RG, since it is equivalent to standard perturbation theory at the $1$-loop level~\cite{Pietroni_2008}.  On a standard $8$-core desktop computer, $1$-loop Time-RG and other $1$-loop computations take $\sim 1$~minute, full Time-RG takes $\sim 1$~hour, and $2$-loop perturbation theories take $\sim 1$~day to compute the power spectrum over $100$ $k$ values in the range $0.001~h/\text{Mpc} \leq k \leq 1~h/\text{Mpc}$.

Figures~\ref{f:P00_M000n0} and~\ref{f:P00_M001n0} show our results for a  massless-neutrino model with cosmological constant (\MoooNo) and an early dark energy model (\MooiNo) whose parameters are given in Table~\ref{t:models}; power spectra have been divided by the ``no-wiggle'' power spectrum $P_\mathrm{nw}(k)$ of Ref.~\cite{Eisenstein_Hu_1998}.  Time-RG is compared to the N-body simulations of Ref.~\cite{Upadhye_2014}, as well as to the $1$- and $2$-loop closure theory calculations of Refs.~\cite{Taruya_2007,Hiramatsu_2009,Taruya_2009} implemented in the~\Copter~code \cite{Carlson_White_Padmanabhan_2009}.  Full and $1$-loop Time-RG are very similar in this range of $k$, with the $1$-loop power spectrum slightly larger at $k \gtrsim 0.25~h/$Mpc.  While $2$-loop closure theory is highly accurate, its computation time makes it difficult to use in an analysis exploring a large parameter space.  Meanwhile, $1$-loop closure theory is comparable to $1$-loop Time-RG in accuracy as well as running time in the $k \leq 0.2~h/$Mpc range.  Although all perturbative calculations break down for sufficiently large $k$, one advantage of Time-RG is that it diverges relatively slowly from N-body calculations, remaining within $\approx 10\%$ of the N-body power spectrum up to $k=0.3~h/$Mpc for $z \geq 0.5$ in both figures.  Figure~\ref{f:hires_M000n0}, a comparison to the high-resolution~\MoooNo~N-body simulation of Ref.~\cite{Upadhye_2014}, shows that both versions of Time-RG are correct in the range $k \leq 1~h/$Mpc to $<10\%$ for $z \geq 1$ and to $<5\%$ for $z \geq 2$.  The figure also shows two other trends: first, that full and $1$-loop Time-RG differ substantially only for $k \gtrsim 1~h/$Mpc; and, second, that full Time-RG underestimates the small-scale power at high $z$ while overestimating it at low $z$.  Figure~\ref{f:Pdv_Pvv_M000n0} compares Time-RG to the velocity power spectrum $P_{\theta\theta}$ and the cross spectrum $P_{\delta\theta}$.  Since the velocity field can only be determined to scales $k \approx 0.5~h/$Mpc given our simulation resolution, we truncate the figure there.  Evidently Time-RG is accurate at the $10\%$ level to $k=0.3~h/$Mpc for both power spectra.  It is not quite as accurate as it was for the density power spectrum, but once again, Time-RG diverges from the N-body spectra smoothly.

\begin{figure}[tp]
  \hskip-3.2in{\Large{(a)}}
  
  \vskip-0.15in
  \includegraphics[width=3.3in]{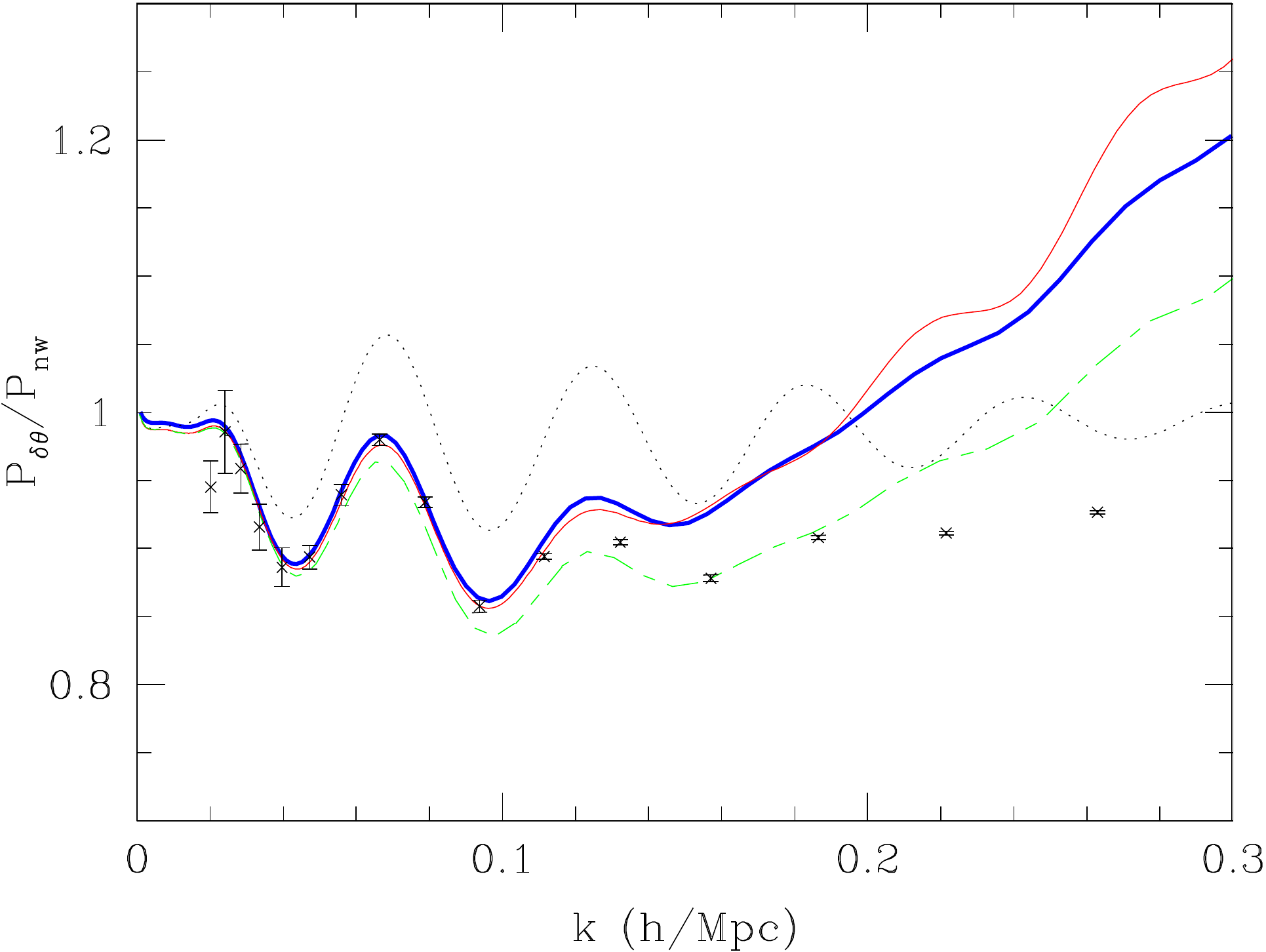}%
  
  \hskip-3.2in{\Large{(b)}}
  
  \vskip-0.15in
  \includegraphics[width=3.3in]{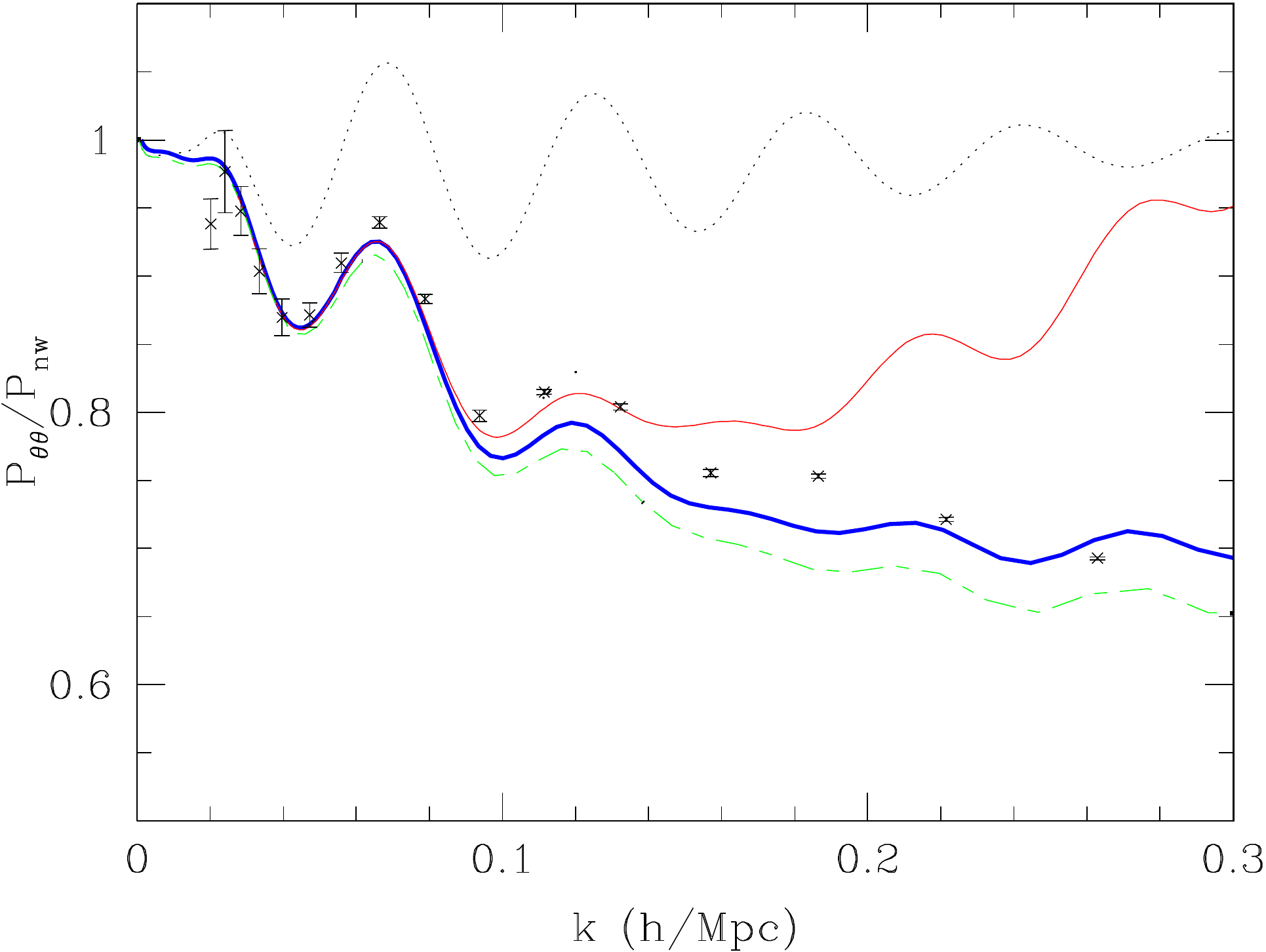}%
  \caption{
    (a) Velocity-density cross-spectrum and (b) velocity auto-spectrum for
    the $\Lambda$CDM model of Ref.~\cite{Carlson_White_Padmanabhan_2009} 
    at $z=0$. Black points are their N-body simulation, and the solid (red) 
    line is $1$-loop Standard Perturbation Theory. Properly including all of the
    $A_{acd,bef}$ integrals of Eq.~(\ref{e:Aacdbef}) in the Time-RG calculation
    leads to an improved fit (solid blue line) at low $k$ relative to
    the Time-RG results of 
    Refs.~\cite{Pietroni_2008,Carlson_White_Padmanabhan_2009}
    (dot-dashed green line).
    Adapted from Figure~8 of Ref.~\cite{Carlson_White_Padmanabhan_2009}.
    \label{f:Pdt_Ptt_fixed}
  }
\end{figure}

We conclude by addressing an error in the original Time-RG algorithm~\cite{Pietroni_2008}.  As noted above, there are $14$ independent components of the $A_{acd,bef}(k)$ integral at each $k$; however, the original work included only $12$ of them, as pointed out in Ref.~\cite{Juergens_2012}.  The erroneous version was implemented in~\Copter, and Ref.~\cite{Carlson_White_Padmanabhan_2009} reported a discrepancy between simulations and the Time-RG predictions of the density-velocity cross-spectrum as well as the velocity auto-spectrum.  In Figure~\ref{f:Pdt_Ptt_fixed}, we add the corrected Time-RG power spectra to their figure.  Including all of the $A_{acd,bef}$ increases $\Pdt$ and $\Ptt$, improving agreement with simulations in the $k < 0.1~h/$Mpc region.  This is significant because the velocity power spectra are important to calculations of the redshift-space power.

\subsection{Massive neutrinos: Linear approximation} 
\label{subsec:massive_neutrinos_linear_approximation}

The coupled system of equations for multiple, non-linear, non-interacting fluids was described in Sec.~\ref{subsec:time-rg_perturbation_theory}.  Since baryonic gas dynamics and feedback effects are only important at small scales, beyond the reach of perturbation theory, we regard the cold dark matter (CDM) and baryons as one single fluid, labelled ``CB.''  We also consider a second fluid ``$\nu$'' consisting of three equal-mass neutrino species, which behave as a warm dark matter component.  Though a generalization to more neutrino species of different masses is straightforward, we focus on the simplest case here.  
Henceforth a subscript ``m'' refers to CDM, baryons, and neutrinos together, so that $\Omegam = \Omega_\mathrm{CDM} + \Omegab + \Omeganu$.
The sum of neutrino masses is related to the neutrino density fraction $\omeganu = \Omega_{\nu,0} h^2$ by $\sum m_\nu = 94 \omeganu$~eV.

Since neutrinos do not cluster strongly, they can further be approximated as a purely linear fluid in Time-RG perturbation theory~\cite{Pietroni_2008,Lesgourgues_etal_2009}.  Thus we may use a linear code such as CAMB or CMBFAST~\cite{Lewis_1999,Seljak_Zaldarriaga_1996,Zaldarriaga_Seljak_Bertschinger_1998} to compute the coupling matrix $\OmegaTRG_\mathrm{CB}$.  Dropping the ``CB'' subscript of $\OmegaTRG$, we have
\begin{eqnarray}
\OmegaTRG_{10}
&\approx&
-\frac{3}{2} \Omegam(\eta) \left[ \fcb + \BetaP(\eta,k) \right]
\\
\BetaP
&=&
\fnu \, \delta_\linnu / \delta_\lincb.
\end{eqnarray}
Using $\delta_\lincb$ rather than $\delta_\mathrm{CB}$ in the denominator of $\BetaP$ makes only a negligible difference~\cite{Lesgourgues_etal_2009}.  Non-linear evolution of $\delta_\nu$ is also not the dominant source of error in this approximation~\cite{Blas_2014,Fuhrer_2015,Castorina_2015}.  Thus Time-RG can incorporate massive neutrinos, or other species leading to scale-dependent CB growth, without much difficulty.

\begin{figure}[tp]
  \includegraphics[width=3.3in]{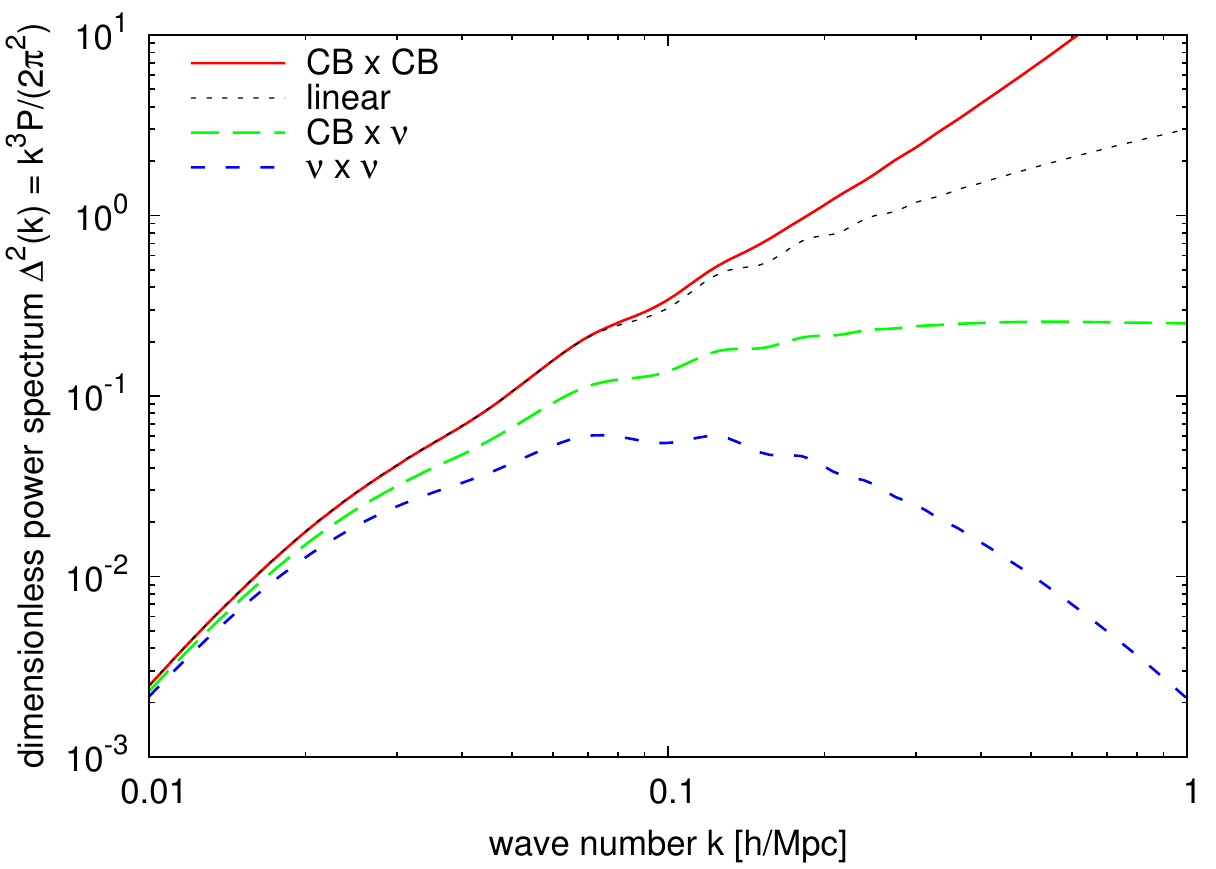}
  \caption{
    Dimensionless power spectra $\Delta^2(k) = \frac{k^3}{2\pi^2} P(k)$ for
    a $\omeganu=0.01$ ($\sum m_\nu=0.94$~eV) model at $z=0$.  The cross-spectrum 
    $P_{\mathrm{CB}\times \nu} \approx \sqrt{\PCB P_\linnu}$ and the
    neutrino auto-spectrum $\Pnu \approx P_\linnu$ use the 
    linear-neutrino approximation.
    \label{f:Delta2}
  }
\end{figure}

In the two-fluid case, the total matter perturbation is $\deltam = (\rhobarcb \deltacb + \rhobarnu \deltanu) / \rhobarm = \fcb \deltacb + \fnu \deltanu$.  In the linear neutrino approximation, the non-linear matter power spectrum is given by
\begin{equation}
\Pmat(k)
\approx
\fcb^2 \PCB
+ 2 \fcb \fnu \sqrt{\PCB P_\linnu}
+ \fnu^2 P_\linnu.
\label{e:lin-nu}
\end{equation}
Figure~\ref{f:Delta2} compares $\PCB$, $P_\linnu$, and $\sqrt{\PCB P_\linnu}$ to the linear CB power spectrum.

Based on the figure, the linear-neutrino approximation should apply even beyond the quasi-linear regime $k \sim 0.1~h/$Mpc.  Non-linear effects in each fluid become important when $\Delta_\mathrm{lin}^2 \sim 1$, as seen by comparing the CB~$\times$~CB and linear curves.  Even for a rather large neutrino mass at $z=0$, the neutrino auto-spectrum remains below $0.1$, while the cross-spectrum plateaus around $0.25$.  If $\Delta_\nu^2(k) - \Delta_\linnu^2(k) \sim   \Delta_\linnu^4(k)$, then the leading-order fractional correction to Eq.~(\ref{e:lin-nu}) will be $\sim (\fnu/\fcb) \sqrt{\Delta^2_\linnu / \Delta^2_\mathrm{CB}} \Delta^2_\linnu$, which is at most $\sim 1\%$.  Moreover, non-linear corrections to $\Delta_\nu^2$ should diminish in importance for $k \gg 0.1~h/$Mpc, allowing a variant of this approximation to be used even in N-body simulations~\cite{Agarwal_2011,Upadhye_2014}.

\subsection{Redshift-space distortions} 

In a perfectly homogeneous universe, there is a precise relation between the redshift $z$ of an object and the comoving distance $\chi(z) = \int_0^z dz' / H(z')$ to that object.  Peculiar velocities $\vec v$ sourced by density inhomogeneities $\delta$ distort this relation, perturbing $\chi$ in the line-of-sight direction $\hat r$ by an amount $\vec v \cdot \hat r / \Hc$ (where $\hat {}$ denotes a unit vector).  If the homogeneous-universe relation is used to identify redshift $z$ with position $\chi(z)$ (``redshift space''), then this peculiar-velocity effect shows up as a direction-dependent distortion of the power spectrum.  Such redshift-space distortions (RSD) enhance the line-of-sight power spectrum in the quasi-linear regime (``squashing'') and suppress it in the non-linear regime (``fingers of god'')~\cite{Jackson_1972}.

A discussion of linear RSD as given in Ref.~\cite{Kaiser_1987} is instructive.  A volume element at comoving position $\vec x$ and redshift $z$, at which point the velocity field takes the value $\vec v(z,\vec x)$, will be assigned a redshift-space position $\vec x_s = \vec x + \hat x \, \vec v \cdot \hat x / \Hc$.  For example, an object falling toward the observer ($\vec v \cdot \hat x < 0$) will appear closer ($|\vec x_s| < |\vec x|$) in redshift space than in real space.  In the flat-sky approximation, $\hat x \approx \hat r = $~constant, and the Jacobian determinant associated with the transformation from real- to redshift-space is $J = |d^3 x / d^3 x_s| = (1 + \partial_x \vec v \cdot r / \Hc)^{-1}$.  

Since the total number of objects (or the total mass) in a given volume does not depend on the coordinates that we use, we can relate the redshift-space overdensity $\delta_s$ to the real-space overdensity by $1+\delta_s = (1+\delta)J$.  In Fourier space, for an irrotational velocity field $\vec v(\vec k) \propto \vec k$, this becomes $\delta_s = (\delta + \muv{k}^2 \theta) J$, where $\muv{k} = \hat k \cdot \hat r$.  In linear theory, $\delta_s \approx (1 + f \muv{k}^2) \delta$, implying the power spectrum
\begin{equation}
P_{s,\mathrm{lin}}(k,\muv{k}) 
=
(1 + f \muv{k}^2)^2 P_\mathrm{lin}(k).
\label{e:Kaiser_lin}
\end{equation}
This qualitatively describes line-of-sight power enhancement expected due to large-scale infall towards matter overdensities, but does not include finger-of-god effects at smaller scales.

In order to extend such a treatment beyond linear theory, two types of corrections have been introduced: streaming models and higher-order corrections to the power spectrum~\cite{Fisher_1995,Kwan_2012}.  The simplest higher-order correction is to replace $\Plin$ in Eq.~(\ref{e:Kaiser_lin}) by the non-linear density power spectrum, while the Scoccimarro anzatz~\cite{Scoccimarro_2004} replaces all three linear power spectra with their non-linear equivalents in order to damp the redshift-space power spectrum:
\begin{equation}
P_{s}(k,\muv{k}) 
=
\Pdd + 2 \muv{k}^2 \Pdt + \muv{k}^4 \Ptt.
\label{e:Scoccimarro}
\end{equation}
All of these  non-linear corrections essentially try to reduce the amplitude of the linear redshift space power spectrum to mimic the effect of non-linear structure formation.

Streaming models explicitly impose finger-of-god suppression by multiplying $P_s(k,\muv{k})$ by a function $\Ffog(f k \sigma_v \muv{k})$, which typically takes a simple form:
\begin{equation}
\Ffog(x)
=
\begin{cases}
  \exp(-x^2)  & \text{ (Gaussian)} \\
  (1+x^2)^{-1} & \text{ (Lorentzian).}
\end{cases}
\label{e:Ffog}
\end{equation}
Here $3\sigma_v^2$ is the trace of the velocity-dispersion tensor.  In practice, it is either fit to the data at each $z$, or approximated by its linear-theory value
\begin{equation}
  \sigma_{v,\mathrm{lin}}^2
  =
  \frac{1}{3} \int \frac{d^3 k}{(2\pi)^3} P_{vv}
  =
  \frac{f^2\Hc^2}{6\pi^2}
  \int dk \, \Plin(k)
  \label{e:sigma_v}
\end{equation}
where $\Plin$ is the linear matter power spectrum.  The convention in the literature is to ``absorb'' the $f^2 \Hc^2$ into the $\sigma_v^2$, so that typical values of $\sigma_{v,\mathrm{lin}} = \sqrt{\int \Plin dk / (6\pi^2)}$ for a $\Lambda$CDM model at $z \sim 1$ are $\sim 1$~Mpc$/h$.

The treatments above approximate the redshift-space power spectrum in terms of the real-space power spectra.  From the form of the Jacobian it is clear that higher-order correlation functions ought to play a role; however, the gradient term in the denominator of $J$ suggests that a simple Taylor expansion will be badly-behaved when $\theta \sim 1$.  Instead, Ref.~\cite{Scoccimarro_2004} uses spatial homogeneity to derive an exact formula for $P_s$ without a badly-behaved Jacobian denominator:
\begin{eqnarray}
P_s(\vec k)
&=&
\int d^3x e^{i \vec k \cdot \vec x}
\left< 
e^{-i \vec k \cdot \hat r  \Delta u} 
\left[\delta(\vec y) + (\hat r \cdot \vec \nabla)u(\vec y)\right] \right.
\nonumber\\
&~&
\qquad
\times
\left.
\left[\delta({\vec y}\,') + (\hat r \cdot \vec \nabla)u({\vec y}\,')\right]
\right>
\label{e:Ps_exact}
\end{eqnarray}
where $u(\vec y) = -\vec v(\vec y) \cdot \hat r / \Hc$, $\Delta u = u(\vec y) - u(\vec y')$, and $\vec x = \vec y - \vec y \, '$.  Recognizing that the Kaiser ``squashing'' and finger-of-god effects are coupled and cannot be treated separately, Taruya, Nishimichi, and Saito (TNS~\TNS) apply a cumulant expansion to the expectation value in Eq.~(\ref{e:Ps_exact}) and find a series of corrections to $P_s$ in terms of higher-order correlation functions.  The leading-order corrections are obtained by neglecting all higher-order correlation functions except for the bispectrum and the disconnected parts of the trispectrum, precisely the approximation used in Time-RG:
\begin{eqnarray}
P_s(k,\mu)
&=&
\Ffog(f \sigma_v k \mu) [
  P_{\delta\delta}(k)
  + 2 \mu^2 P_{\delta\theta}(k)
  + \mu^4 P_{\theta\theta}(k)
\nonumber\\&&
  \qquad \qquad \qquad
  + \Pbis(k,\mu)
  + \Ptri(k,\mu)
]
\label{e:Ps_TNS}
\end{eqnarray}
\begin{eqnarray}
\frac{\Pbis(k,\mu)}{k\mu}
&=&
\int \frac{d^3q}{(2\pi)^3} \frac{\muv{q}}{q}
\Big[
  \Btns(\vec q,\vec k - \vec q, -\vec k)
\nonumber\\
&&
  \quad - \Btns(\vec q, \vec k, -\vec k - \vec q)
\Big]
\label{e:Pbis_tns}
\\
\frac{\Ptri(k,\mu)}{k^2 \mu^2}
&=&
\int \frac{d^3q}{(2\pi)^3} \ttns(\vec q) \ttns(\vec k - \vec q)
\label{e:Ptri_tns}
\\
\Btns(\vec k_1, \vec k_2, \vec k_3)
&=&
\Btdd(\vec k_1, \vec k_2, \vec k_3)
- \muv{k_2}^2 \Bttd(\vec k_1, \vec k_2, \vec k_3)
\nonumber\\
&&
- \muv{k_3}^2 \Btdt(\vec k_1, \vec k_2, \vec k_3)
\nonumber\\
&&
+ \muv{k_2}^2 \muv{k_3}^2 \Bttt(\vec k_1, \vec k_2, \vec k_3)
\label{e:Btns}
\\
\ttns(\vec k_1)
&=&
\frac{\muv{k_1}}{k_1}
\left[
  \Pdt(\vec k_1)
  + \muv{k_1}^2 \Ptt(\vec k_1)
\right]
\label{e:ttns}
\end{eqnarray}
Here and henceforth, $\mu$ is assumed to mean $\muv{k}$ unless otherwise labelled.  A thorough study of approximations to $P_s$ finds this approach to be the most successful at matching N-body calculations and at providing an unbiased estimate of the growth rate $f$~\cite{Kwan_2012}.  

In practice, Ref.~\TNS~computes the power spectra in Eq.~(\ref{e:Ps_TNS}) using closure theory, the bispectra in Eq.~(\ref{e:Pbis_tns}) using the tree-level approximation of Ref.~\cite{Fry_1984}, and the power spectra in Eq.~(\ref{e:Ptri_tns}) using linear theory.  In the next Section, we compute all of these terms together in Time-RG perturbation theory, making only the standard Time-RG approximation of neglecting the connected trispectrum.  There are three potential advantages to this approach:
\begin{enumerate}
\item Time-RG is easily-generalized to multi-component models with scale-dependent growth; 
\item Time-RG is well-behaved at $k \lesssim 1~h/$Mpc for $z \gtrsim 1$, with errors $\lesssim 10\%$; and
\item errors may be estimated by approximating the connected trispectrum, as in Ref.~\cite{Juergens_2012}.
\end{enumerate}
Regarding the second of these, the most promising perturbative RSD treatments employ a streaming function $\Ffog$, a phenomenological model which is not expected to be correct beyond the $\approx 5\%$ level at quasi-linear scales $k \gtrsim 0.1~h/$Mpc.  Thus percent-level accuracy in the computation of the power spectra is unnecessary; much more important is the calculation of $P(k)$ at the $5\% - 10\%$ level over a larger range of scales.

\section{Redshift-space distortions in Time-RG perturbation theory}
\label{sec:rsd_in_trg}

\subsection{$\Pbis$ and $\Ptri$ in Time-RG} 

Further analysis can substantially simplify the computation of $\Pbis$ and $\Ptri$.  Though we defer the details to Appendices~\ref{sec:evolution_of_Pbis} and~\ref{sec:computation_of_Ptri}, we summarize the results here. First, the $\mu$-dependence of $P_s(k,\mu)$ can be separated out, yielding a polynomial in $\mu^2$ with $k$-dependent coefficients:
\begin{eqnarray}
  P_s(k,\mu)
  &=&
  \Ffog(f\sigma_v k \mu) 
  [ P_0(k) + P_2(k)\mu^2 + P_4(k)\mu^4 
    \nonumber\\
    &&
    \qquad\qquad\qquad + P_6(k)\mu^6 + P_8(k)\mu^8 ].
  \label{e:Pj}
\end{eqnarray}
Both $\Pbis$ and $\Ptri$ contribute to $P_2$, $P_4$, and $P_6$, while $\Ptri$ also contributes to $P_8$:
\begin{eqnarray}
\Pbis(k,\mu)
&=&
\Pbisj{2}(k)\mu^2 
+ \Pbisj{4}(k)\mu^4 \
+  \Pbisj{6}(k)\mu^6
\\
\Ptri(k,\mu)
&=&
\Ptrij{2}(k)\mu^2
+ \Ptrij{4}(k)\mu^4
\nonumber\\
&&
+ \Ptrij{6}(k)\mu^6
+ \Ptrij{8}(k)\mu^8.
\end{eqnarray}

\begin{figure}[tp]
  \begin{center}
    \includegraphics[width=3.3in]{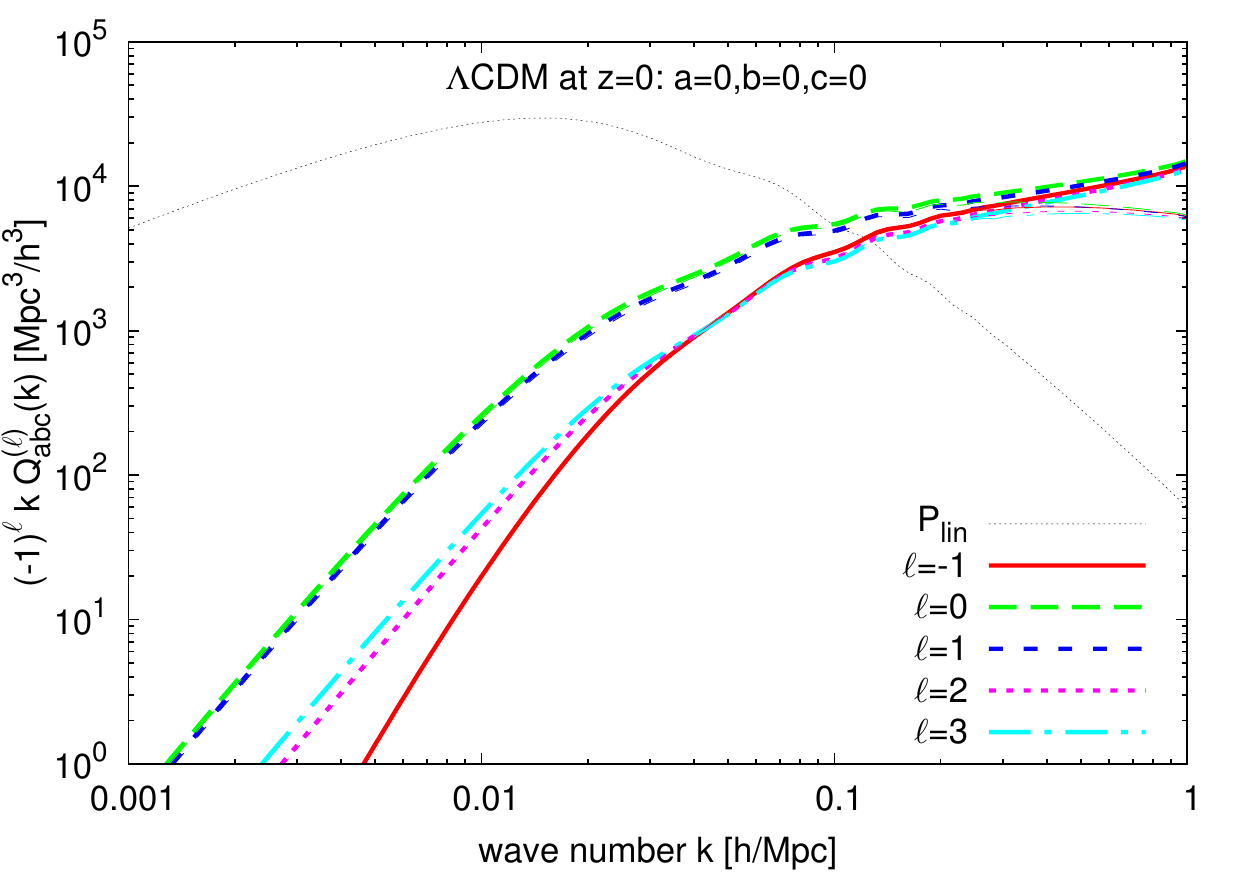}
    \caption{
      Bispectrum integrals $(-1)^\ell k \, \Qlabc(k)$ 
      with $a=b=c=\delta$, for~\MoooNo~at 
      $z=0$.  Thick lines show full Time-RG, while thin lines show 
      its $1$-loop approximation.  
      The other components are similar in magnitude as well as 
      $k$-dependence.  The linear power spectrum is included for comparison.
      \label{f:Ql000}
    }
  \end{center}
\end{figure}

Second, rather than keeping track of the full functional dependence of the bispectrum in order to integrate Eq.~(\ref{e:Pbis_tns}) for the $\Pbisj{j}$, we can break up the bispectrum integrals into a series of terms $\Qlabc$, with $-1 \leq \ell \leq 3$, which depend only on $k$ and $\eta$.  In the Time-RG framework, the evolution equations for the $\Qlabc$ follow from Eq.~(\ref{e:eom_full_B}).  Thus the $\Qlabc$ are analogous to the $I_{acd,bef}(k)$ in real-space Time-RG~(\ref{e:Iacdbef}).  Appendix~\ref{sec:evolution_of_Pbis} expresses the $\Pbisj{j}$ as linear combinations of $k \Qlabc$ and then derives these evolution equations, which require the computation of a two-dimensional integral analogous to $A_{acd,bef}$ at each time step.  Figure~\ref{f:Ql000} compares $k \Qlabc$ to the linear power spectrum, showing that non-linear contributions to redshift-space distortions are non-negligible at the BAO scale.  The figure also shows $1$-loop approximations to $\Qlabc$.

Meanwhile, computation of the $\Ptrij{j}$ is straightforward.  We describe it in Appendix~\ref{sec:computation_of_Ptri} for completeness, but the only change relative to Ref.~\TNS~is that we carry out the computation using the non-linear power spectra.

  The full set of evolution equations for redshift-space distortions in Time-RG perturbation theory is then Eqs.~(\ref{e:eom_P},\ref{e:Iacdbef},\ref{e:eom_I},\ref{e:Aacdbef},\ref{e:eom_Q},\ref{e:Rlabc}).  Since the bispectra are small at early times, the $\Qlabc$ can be initialized to zero, as with the $I_{acd,bef}$.  Reference~\cite{Blas_2014} finds that the resulting error is approximately $1/\zin$ for an initial redshift $\zin$, and our numerical results with $\zin=200$ are consistent with this.  Redshift-space Time-RG follows the $40$ terms $\Qlabc$ as well as $14$ unique $I_{acd,bef}$, so we can expect a fourfold increase in computation time relative to real-space Time-RG.  Our implementation of redshift-space Time-RG, {\tt{redTime}}, is available on-line at \url{http://www.hep.anl.gov/cosmology/pert.html}.  {\tt{redTime}} uses the GNU Scientific Library~\cite{Galassi_2009} to evolve the equations of motion, and the CUBA Library~\cite{Hahn_2005} to compute numerically the multi-dimensional integrals in Eqs.~(\ref{e:Aacdbef},\ref{e:Rlabc},\ref{e:Ptrij}).  The redshift-space and multipole plots in this article have been produced using CAMB for transfer functions and {\tt{redTime}} for Time-RG calculations.

\subsection{Comparison to the literature} 

\begin{figure}[tp]
  \hskip-3.2in{\Large{(a)}}
  
  \vskip-0.25in
  \includegraphics[width=3.3in]{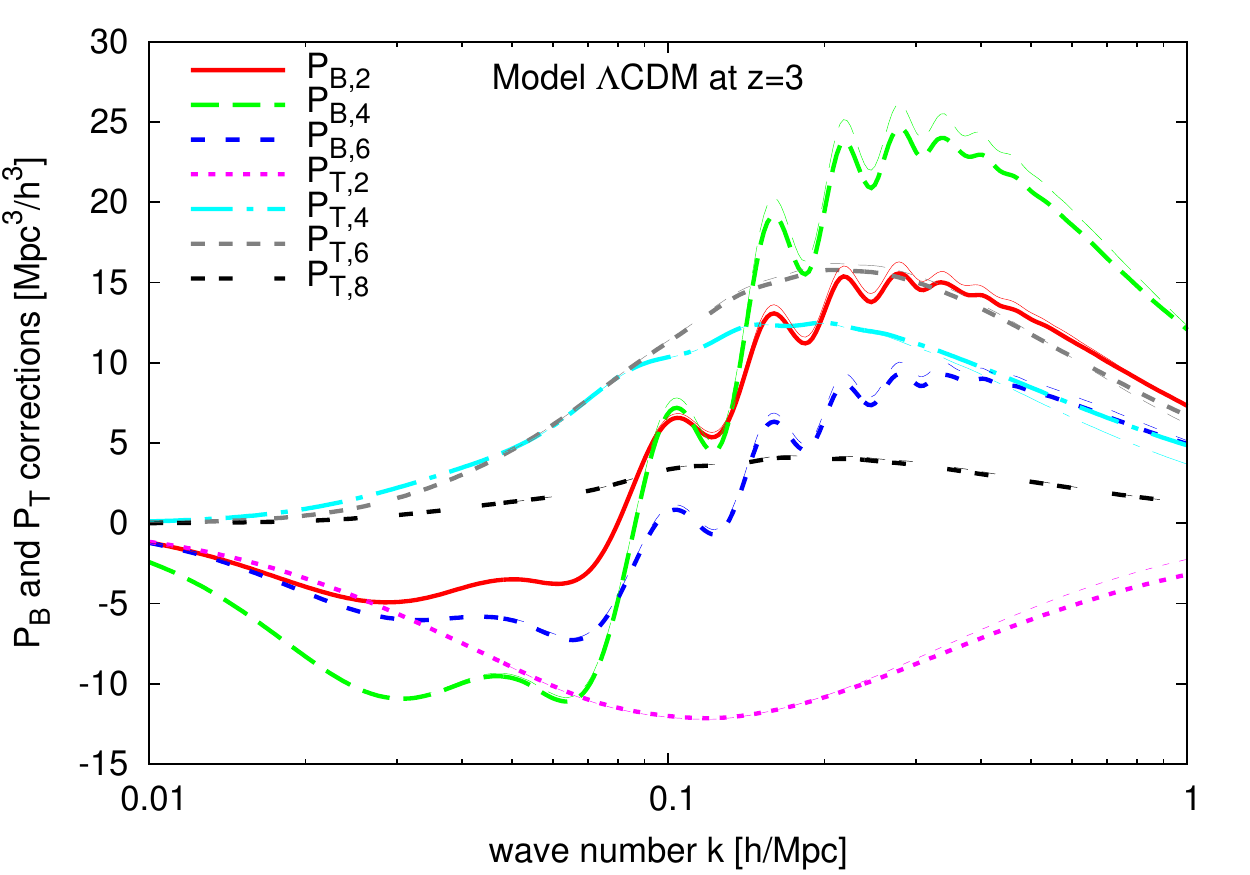}%
  
  \hskip-3.2in{\Large{(b)}}
  
  \vskip-0.25in
  \includegraphics[width=3.3in]{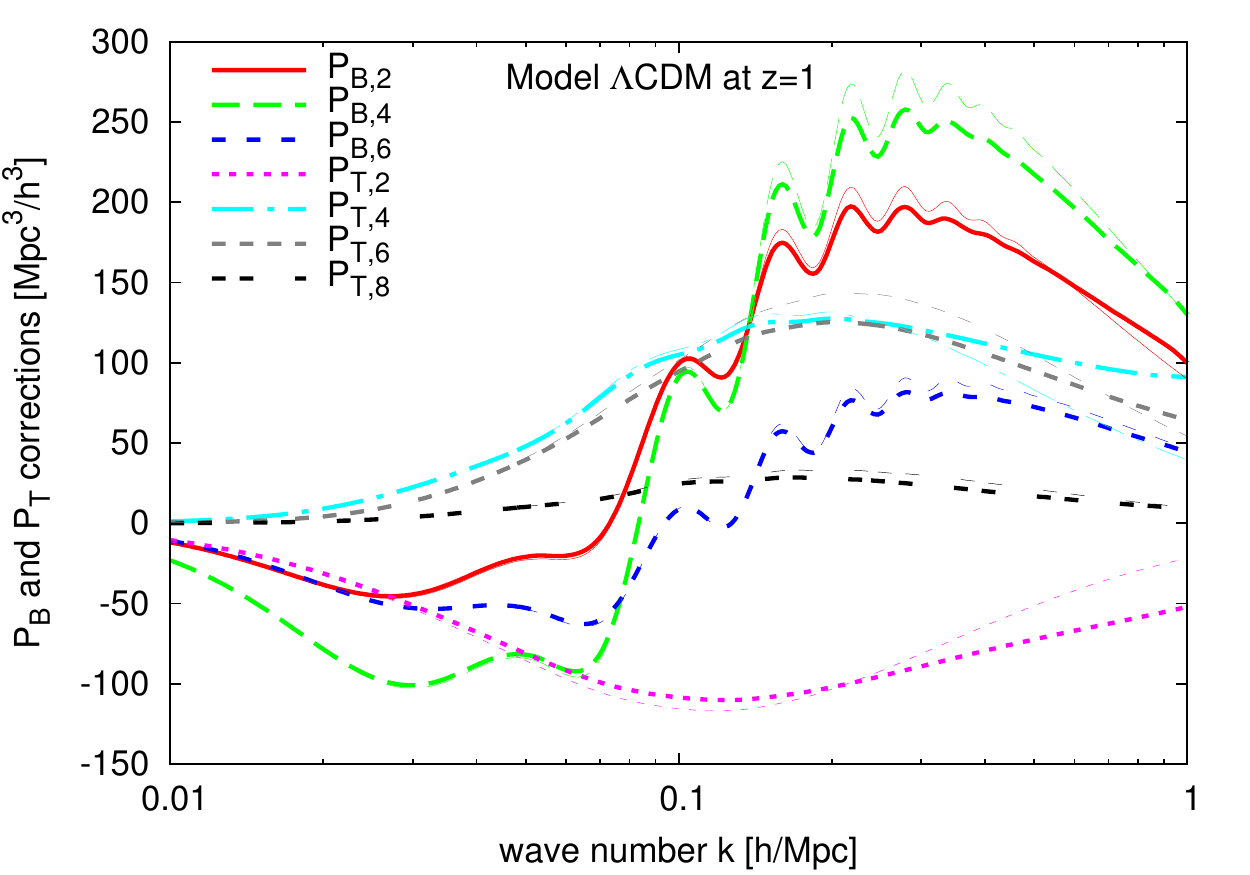}%

  \hskip-3.2in{\Large{(c)}}

  \vskip-0.25in
  \includegraphics[width=3.3in]{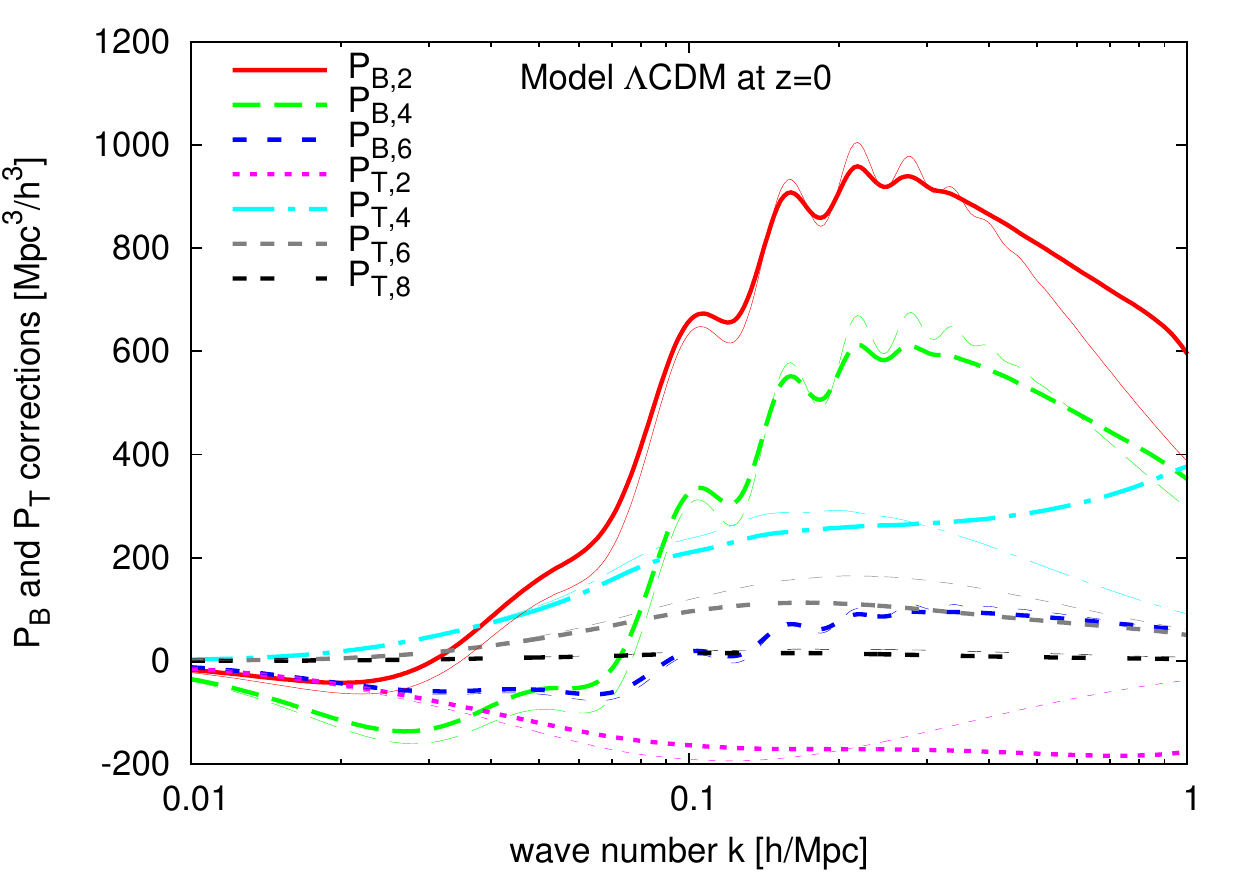}%

  \caption{
    Non-linear corrections $P_{B,j}(k)$ and $P_{T,j}(k)$ to the redshift-space
    power spectrum.  Time-RG calculations are shown as thick lines, and 
    the results of~\TNS~are thin lines.~\MoooNo~is 
    shown at: (a) $z=3$; (b) $z=1$; (c) $z=0$.
    \label{f:trg_vs_tns_corr}
  }
\end{figure}

\begin{figure}[tp]
  \hskip-3.2in{\Large{(a)}}
  
  \vskip-0.25in
  \includegraphics[width=3.3in]{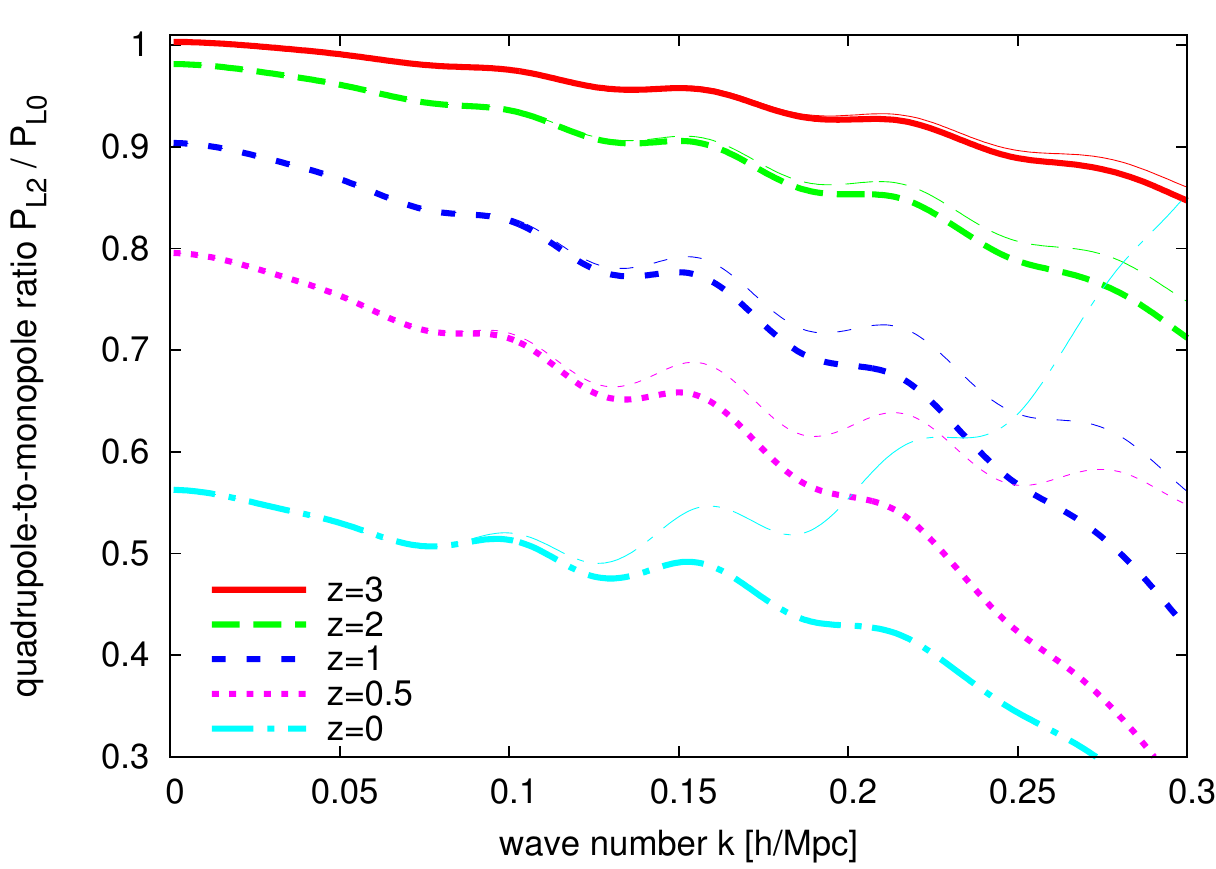}%
  
  \hskip-3.2in{\Large{(b)}}
  
  \vskip-0.2in
  \includegraphics[width=3.3in]{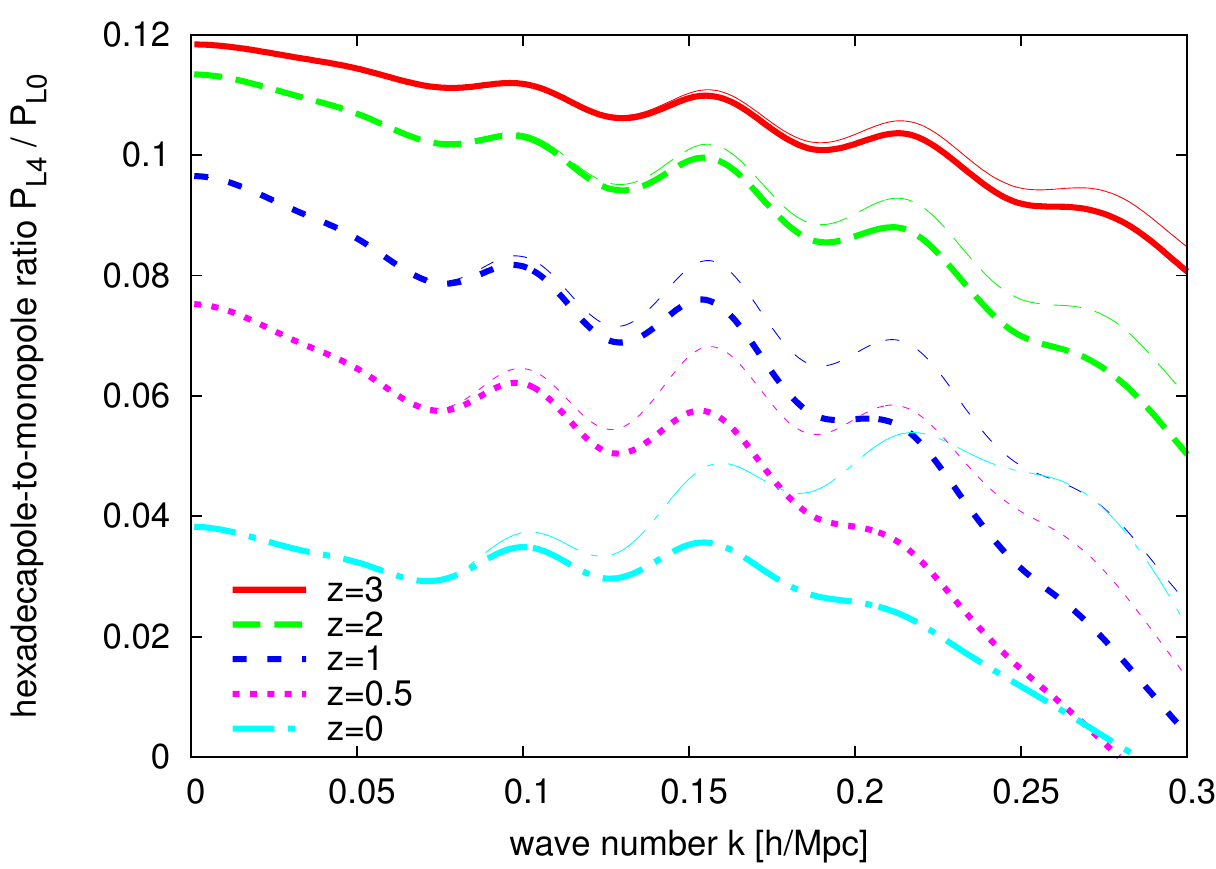}%
  \caption{
    Ratios of the (a) quadrupole and (b) hexadecapole to the monopole
    of the redshift-space power spectrum $P_s(k,\mu)$, shown for the
    model~\MoooNo.  Time-RG calculations are shown as thick
    lines and TNS calculations as thin lines, with $\sigma_v$ taken
    from linear theory.
    \label{f:trg_vs_tns_quad_hexa}
  }
\end{figure}

Figures~\ref{f:trg_vs_tns_corr} and~\ref{f:trg_vs_tns_quad_hexa} compare our calculations to those of Ref.~\TNS~for a $\Lambda$CDM cosmology.  The correction terms $\Pbisj{j}$ and $\Ptrij{j}$ are directly compared in Fig.~\ref{f:trg_vs_tns_corr}, which shows that the two sets of results agree closely at early times and large scales.  Ref.~\TNS~predicts somewhat larger corrections at $k \leq 0.2~h/$Mpc, while Time-RG corrections become larger in magnitude around $k \sim 1~h/$Mpc.

Multipole moments of the redshift-space power spectrum $P_s(k,\mu)$ are found by projecting Eq.~(\ref{e:Ps_TNS}) onto a basis of Legendre polynomials $\Pleg_\ell(\mu)$ for even $\ell$:
\begin{equation}
\PLl(k) 
=
\frac{2\ell + 1}{2}
\int_{-1}^1 d\mu \, P_s(k,\mu) \Pleg_\ell(\mu).
\label{e:multipoles}
\end{equation}
Defining $\PLl = \sum_{j=0} {\mathscr M}_{\ell,2j} P_{2j}$, $m_n = \int_{-1}^1 d\mu \, \mu^{2n} \Ffog$, $\alpha = f k \sigma_v$, and the Legendre coefficients $p_{\ell i}$ such that $\Pleg_\ell(x) = \sum_i p_{\ell i} x^{2i}$ for even $\ell$, we have ${\mathscr M}_{\ell,2j} = \frac{2\ell+1}{2} \sum_i p_{\ell,i} m_{i+j}$.  For a Gaussian $\Ffog(\alpha\mu)$, $m_n^\mathrm{G} =  \gamma(\frac{2n+1}{2},\alpha^2) / \alpha^{2n+1}$, where $\gamma(a,x) = \int_0^x e^{-t} t^{a-1} dt$ is the incomplete gamma function.  For a Lorentzian $\Ffog$, $m_0^\mathrm{L} = 2 \alpha^{-1} \arctan(\alpha)$, and for $n \geq 1$ we have the recursion relation  $\alpha^2 m_n^\mathrm{L} = 2/(2n-1) - m_{n-1}^\mathrm{L}$.

Ratios of the quadrupole ($\ell=2$) and hexadecapole ($\ell=4$) to the monopole ($\ell=0$) are shown in Fig.~\ref{f:trg_vs_tns_quad_hexa} for the Time-RG and TNS~\TNS~calculations.  At early times and large scales, the quadrupole and hexadecapole ratios approach their linear Einstein-de Sitter values $50/49$ and $6/49$, respectively.  At smaller scales, Time-RG and TNS agree at early times, while TNS predicts substantially higher quadrupole-to-monopole and hexadecapole-to-monopole ratios at late times.  This is due in part to the fact that Time-RG overestimates the late-time non-linear power spectrum $P_{\delta\delta}$, while the closure theory calculation of Ref.~\TNS~underestimates it~\cite{Carlson_White_Padmanabhan_2009}.

\subsection{Redshift-space power spectrum} 

The two main aims of this article are to compute the effects of massive neutrinos and evolving dark energy on the redshift-space power spectrum $P_s(k_\parallel,k_\perp)$, and to demonstrate that our results are consistent with N-body simulations.  We are now in a position to do the first of these. 

\begin{figure}[tp]
  \includegraphics[width=3.4in]{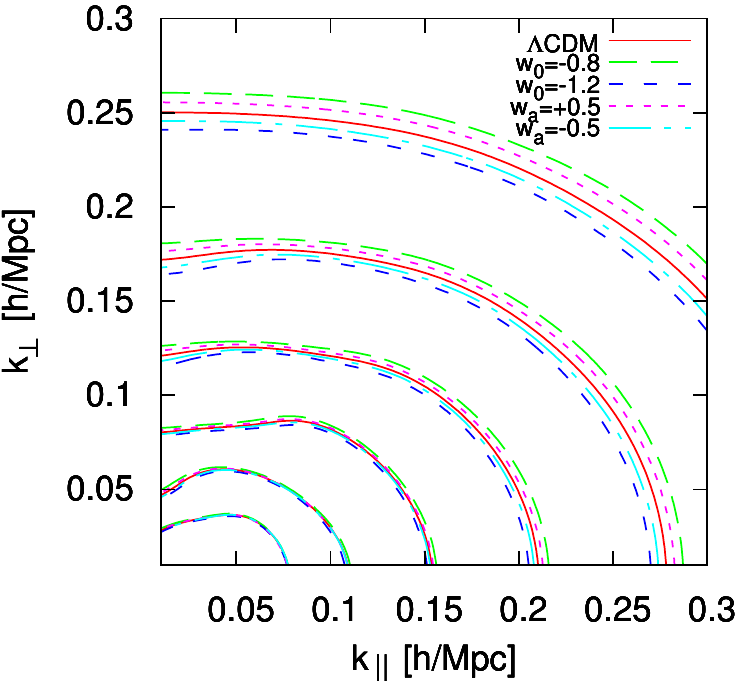}%
  \caption{
    Logarithmic contours of the redshift-space power spectrum 
    $P_s(k_\parallel,k_\perp)$ at $z=1$.  
    The equation of state parameters $w_0$ and $w_a$
    are varied from their~\MoooNo~values $w_0=-1$, $w_a=0$, with all
    other parameters held fixed.
    \label{f:Ps_vary_w0wa}
  }
\end{figure}

Figure~\ref{f:Ps_vary_w0wa} shows the redshift-space power spectrum for a range of dark energy models.  As either $w_0$ or $w_a$ is increased, leading to an overall increase in $w(z)$, the power spectrum increases both parallel and perpendicular to the line of sight.  Effectively the perpendicular direction constrains the growth factor $D(z)$ while the $\mu$-dependence of $P_s$ constrains the derivative $f = -d \log D/d \log (1+z)$.

\begin{figure}[tp]
  \includegraphics[width=3.4in]{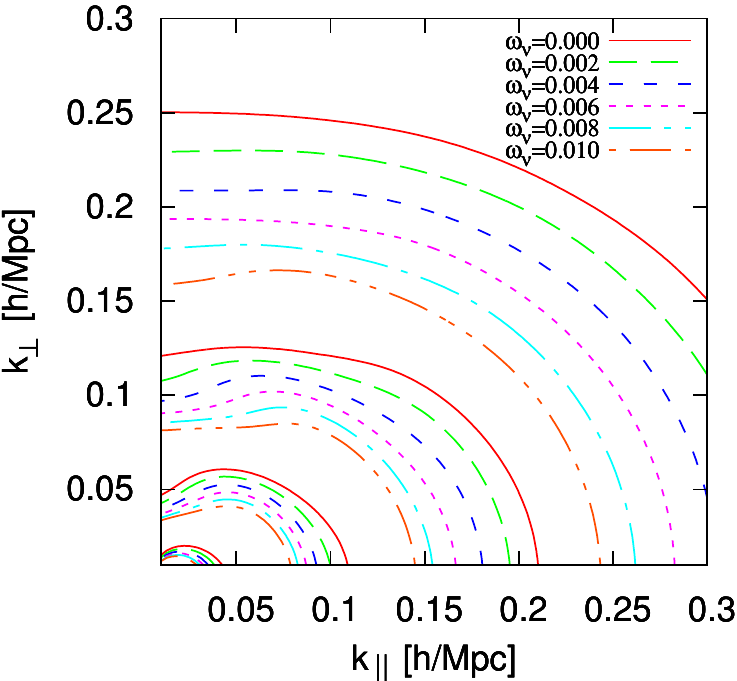}%
  \caption{
    Logarithmic contours of the redshift-space power spectrum 
    $P_s(k_\parallel,k_\perp)$ at $z=1$.  The neutrino density parameter
    $\omega_\nu = \Omega_{\nu,0} h^2$ is varied from its~\MoooNo~value of zero,
    with the early-time (large-scale) power spectrum normalization equal
    for all $\omega_\nu$, and all other parameters fixed at their~\MoooNo~
    values.  This normalization convention corresponds to $\sigma_8$ of
    $0.8$, $0.754$, $0.705$, $0.660$, $0.618$, and $0.580$ for $\omega_\nu$
    of $0$, $0.002$, $0.004$, $0.006$, $0.008$, and $0.01$, respectively.
    \label{f:Ps_vary_wnu}
  }
\end{figure}

\begin{figure}[tp]
  \hskip-3.2in{\Large{(a)}}
  
  \vskip-0.25in
  \includegraphics[width=3.3in]{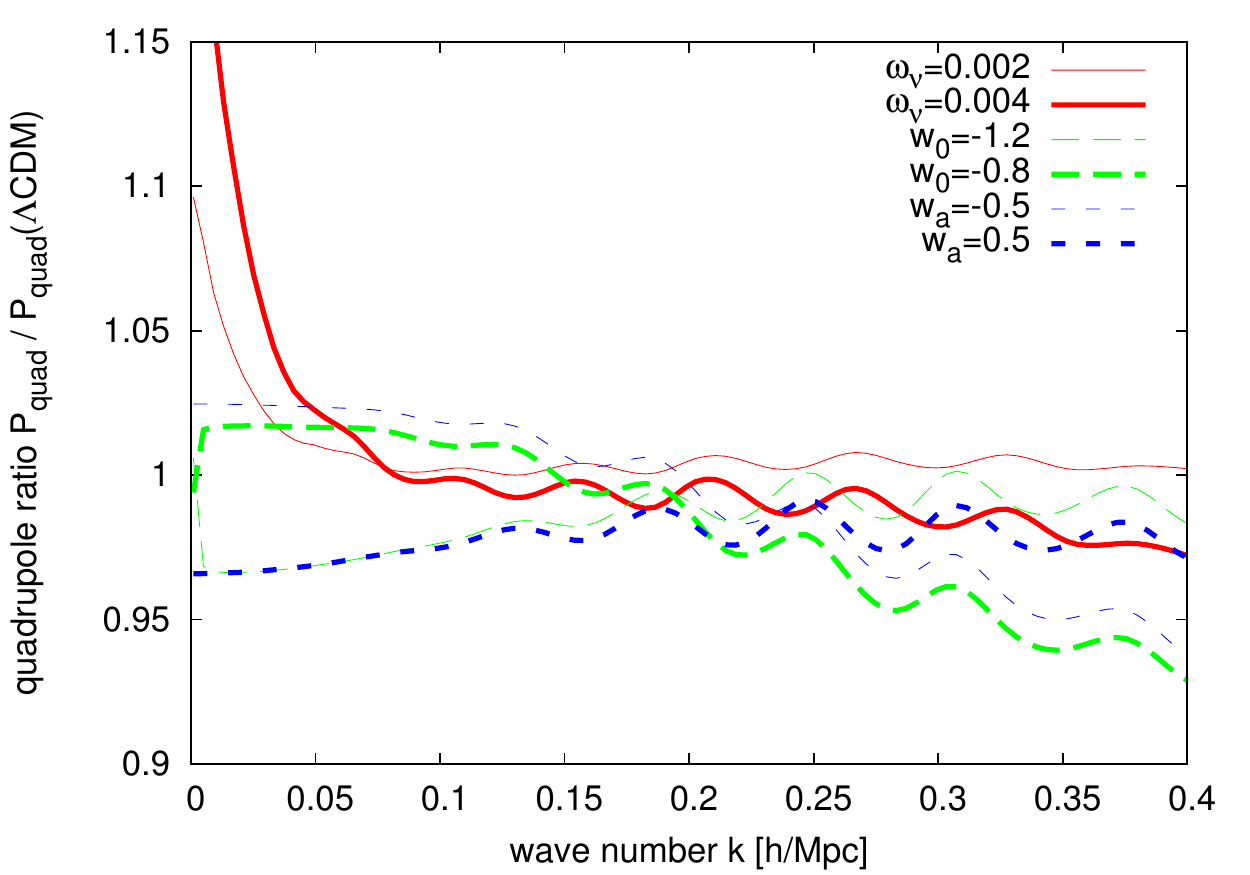}%
  
  \hskip-3.2in{\Large{(b)}}
  
  \vskip-0.2in
  \includegraphics[width=3.3in]{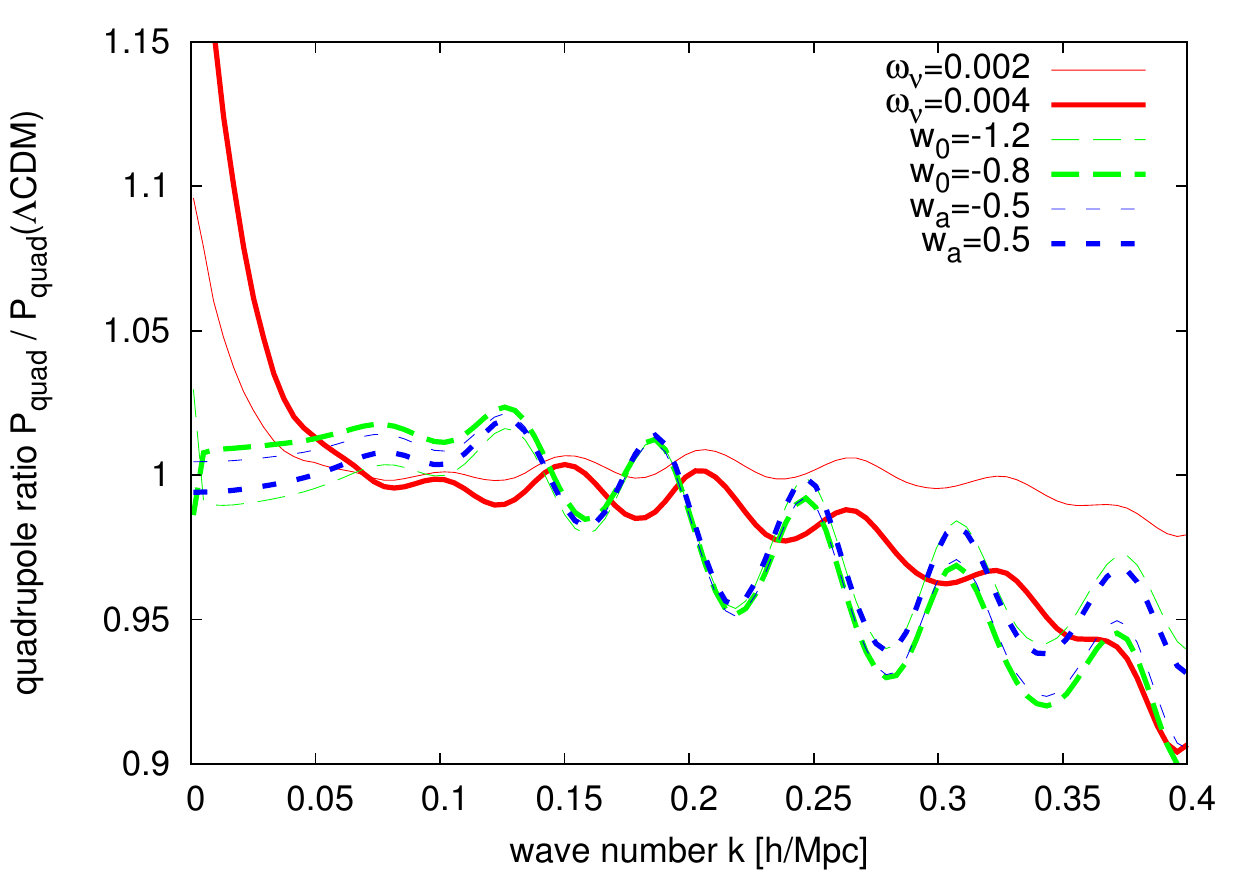}%
  \caption{
    Sensitivity of the quadrupole to changes in $\omega_\nu$, $w_0$, and $w_a$
    relative to their values in model~\MoooNo~at (a)~$z=1$ and (b)~$z=0$. 
    Shown for each model is the ratio of its quadrupole to that 
    of~\MoooNo, assuming linear theory for $\sigma_v$.
    \label{f:Pquad_varyw0wanu}
  }
\end{figure}

Meanwhile, Figure~\ref{f:Ps_vary_wnu} varies the neutrino density fraction in $\Lambda$CDM models while keeping the early-time power spectrum normalization constant.  The resulting variation in $\sigma_8$, the late-time normalization, is the largest contributor to the differences among the power spectra in Fig.~\ref{f:Ps_vary_wnu}.  Since Cosmic Microwave Background (CMB) measurements constrain the early-time normalization, neutrino constraints are strongest when CMB and large-scale structure data are combined~\cite{Font-Ribera_2014}.  

Aside from differences in $\sigma_8$, the models in Fig.~\ref{f:Ps_vary_wnu} differ in their redshift-space dependence to a much greater extent than those in Fig.~\ref{f:Ps_vary_w0wa}.  Figure~\ref{f:Pquad_varyw0wanu} presents this effect in a different way by showing the fractional changes in the quadrupole caused by varying $\omega_\nu$, $w_0$, and $w_a$ relative to their massless-neutrino~\MoooNo~values.
This dependence arises from the small-scale suppression of growth by massive neutrinos, which enters the linear redshift-space power spectrum through a scale-dependence in $f(z,k) = -\partial \log D(z,k) / \partial \log (1+z)$, a suppression of non-linear corrections, and a greater linear-theory velocity dispersion.  Thus an analysis of redshift-space distortions provides one more method for distinguising between the cosmological effects of massive neutrinos and evolving dark energy.

\section{Comparison with simulations}
\label{sec:comparison_with_simulations}

\subsection{N-body simulations with HACC} 
\label{subsec:n-body_simulations_with_hacc}

In order to test the accuracy of our perturbative calculation, we have run a suite of N-body cosmological simulations using the Hardware/Hybrid-Accelerated Cosmology Code (HACC), described in Ref.~\cite{Habib_2014}.  For each model in Table~\ref{t:models} we ran a high-resolution TreePM simulation with $3200^3$ particles and a box size of $(2.1~\text{Gpc})^3$.  Additionally, we averaged the results of $16$ PM runs, each with $512^3$ particles and a box size of $(1.3~\text{Gpc})^3$, in order to reduce the simulation error at quasi-linear scales $k \sim 0.1~h/$Mpc.  Redshift-space power spectra $P_s(k_\parallel, k_\perp)$ were computed in the distant-observer approximation.  Assuming the observer to be located far from the simulation volume along one of the three coordinate axes, we used each particle's line-of-sight velocity to shift its position from real space to redshift space.  Fourier transformation of this shifted particle distribution yields a measurement of $P_s$, which can then be averaged over multiple simulation runs and all three lines of sight.

Our simulations treat massive neutrinos using the linear approximation of Refs.~\cite{Saito_2008,Agarwal_2011}, as described in Sec.~\ref{subsec:massive_neutrinos_linear_approximation}.  Particles in the simulation represent only the baryons and cold dark matter.  After their non-linear power spectrum has been computed, the linear neutrino power (calculated using Boltzmann integrators such as CAMB and CMBFAST~\cite{Lewis_1999,Seljak_Zaldarriaga_1996,Zaldarriaga_Seljak_Bertschinger_1998,Zaldarriaga_Seljak_2000}) is added using Eq.~(\ref{e:lin-nu}).  This approximation was shown in Ref.~\cite{Upadhye_2014} to agree well with perturbative results, with a discrepancy at large $z$ and $\omega_\nu$ that can approximately be corrected by including the scale-dependence of the linear growth factor.  For an alternative approach to simulating massive neutrino cosmologies, see, for example,~\cite{Castorina_2015}.

\subsection{$m_\nu=0$ $\Lambda$CDM and $\sigma_v$ fitting} 
\label{subsec:lcdm_and_sigv_fitting}

\begin{figure}[tp]
  \includegraphics[width=3.4in]{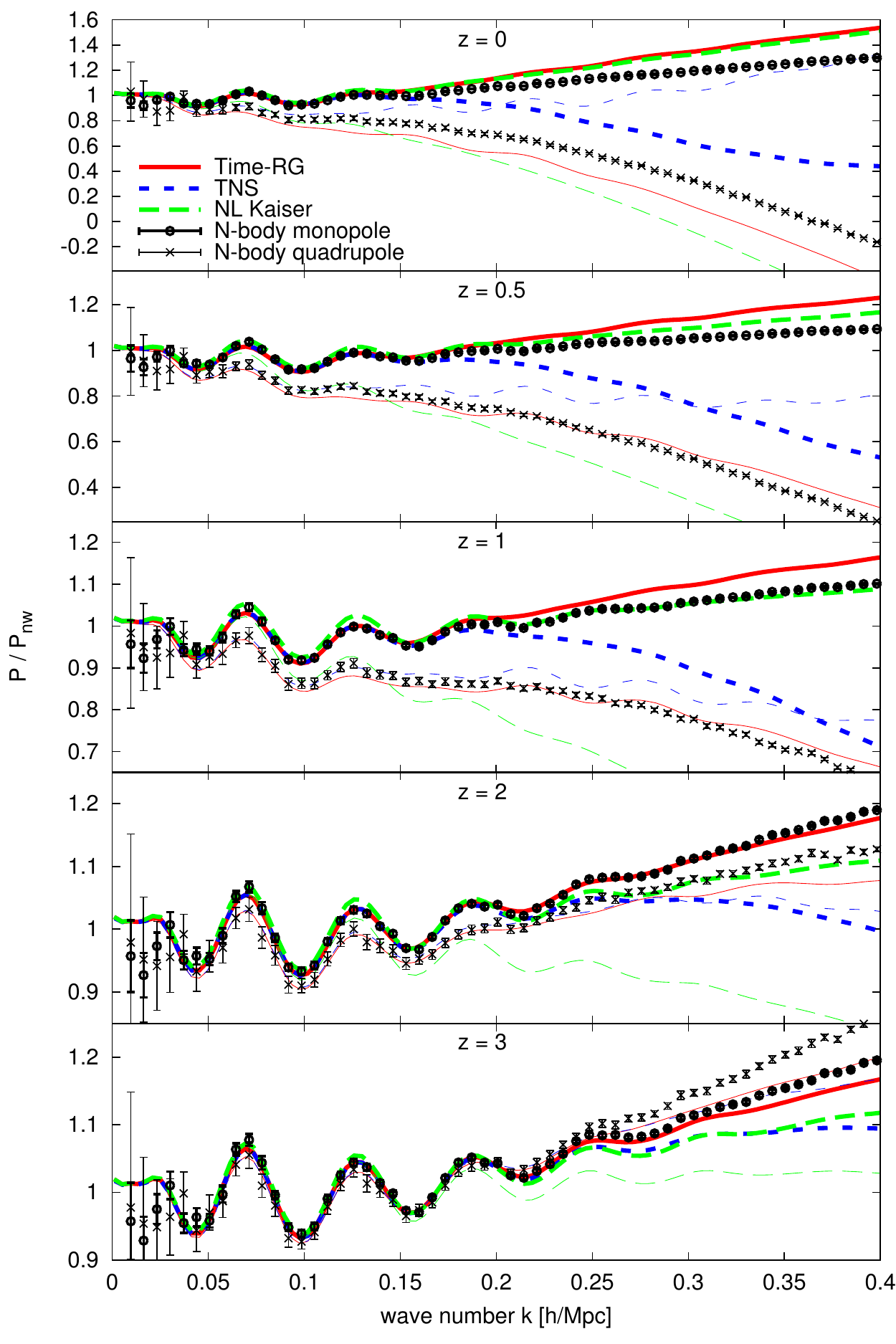}
  \caption{
    Monopoles (thick lines) and quadrupoles (thin lines) of $P_s(k,\mu)$ 
    computed using Time-RG, TNS~\TNS, non-linear Kaiser, 
    and N-body methods, for the $m_\nu=0$ model~\MoooNo.
    \label{f:multipoles_M000n0}
  }
\end{figure}

Perturbative calculations for the massless-neutrino model~\MoooNo~are compared with N-body simulations in Figure~\ref{f:multipoles_M000n0}.  At each redshift, the velocity dispersion parameter $\sigma_v$ in the Lorentzian streaming function is fit to the N-body monopole and quadrupole by minimizing $\chi^2$.  The N-body error in each $k$ bin is the quadrature sum of the sample variance and the run-to-run standard deviation in the $16$ simulation runs, divided by $\sqrt{16}$.  Since errors in the perturbative calculations in the fully non-linear regime would bias the $\sigma_v$ fitting, resulting in a worse fit at low $k$, we restricted the fitting procedure to the range $k < k_{\max} = 0.2~h/$Mpc.  Even then, at $z=0$, errors in the perturbative calculations appear to bias $\sigma_v$.  Time-RG in real space overestimates $P(k)$ at $k \sim 0.2~h/$Mpc, leading to an overestimate of $\sigma_v$, hence an underestimate in the quadrupole. Closure theory at $1$-loop, used in the TNS calculation, has the opposite effect.  For $z \gtrsim 1$, both perturbative calculations agree well with the simulation.  The $\sigma_v$ fitting procedure is discussed further in Appendix~\ref{sec:error_bounds_and_sigv_fits}, which shows that our results are insensitive to the value of $k_{\max}$ within the range $0.15~h/\mathrm{Mpc} \leq k_{\max} \leq 0.25~h/$Mpc. 

\begin{table}
  \begin{center}
    \caption{Wave number $k$ [$h/$Mpc] below which each perturbation theory
      is accurate to the given accuracy level (Acc.) in 
      the~\MoooNo~massless-neutrino model.  ``NL Kaiser'' uses the Time-RG
      real-space power spectrum in the Kaiser RSD 
      formula, Eq.~(\ref{e:Kaiser_lin}).
      \label{t:PTerrorM000n0}} 
    \tabcolsep=0.2cm
    \begin{tabular}{lc|cc|cc|cc}
      $z$   & Acc.   &\multicolumn{2}{c}{Time-RG}&\multicolumn{2}{c}{TNS~\TNS}&\multicolumn{2}{c}{NL Kaiser}\\
            &        & $\ell=0$    & $\ell=2$    & $\ell=0$     & $\ell=2$    & $\ell=0$     & $\ell=2$\\
      \hline
      $0$   & $1\%$  & $0.13$      & $0.09$      & $0.14$       & $0.11$      & $0.08$       & $0.12$  \\
            & $2\%$  & $0.13$      & $0.09$      & $0.15$       & $0.11$      & $0.08$       & $0.13$  \\
            & $5\%$  & $0.18$      & $0.11$      & $0.17$       & $0.13$      & $0.17$       & $0.13$  \\
            & $10\%$ & $0.26$      & $0.11$      & $0.18$       & $0.14$      & $0.27$       & $0.14$  \\
      \hline
      $0.5$ & $1\%$  & $0.15$      & $0.11$      & $0.17$       & $0.15$      & $0.11$       & $0.07$  \\
            & $2\%$  & $0.19$      & $0.11$      & $0.17$       & $0.15$      & $0.12$       & $0.07$  \\
            & $5\%$  & $0.21$      & $0.29$      & $0.19$       & $0.15$      & $0.32$       & $0.15$  \\
            & $10\%$ & $0.33$      & $0.33$      & $0.23$       & $0.20$      & $1.17$       & $0.19$  \\
      \hline
      $1$   & $1\%$  & $0.19$      & $0.28$      & $0.19$       & $0.16$      & $0.08$       & $0.07$  \\
            & $2\%$  & $0.20$      & $0.33$      & $0.19$       & $0.20$      & $0.12$       & $0.07$  \\
            & $5\%$  & $0.37$      & $0.40$      & $0.23$       & $0.21$      & $0.70$       & $0.19$  \\
            & $10\%$ & $1.07$      & $0.47$      & $0.28$       & $0.32$      & $0.92$       & $0.20$  \\
      \hline
      $2$   & $1\%$  & $0.36$      & $0.24$      & $0.21$       & $0.28$      & $0.21$       & $0.19$  \\
            & $2\%$  & $0.43$      & $0.30$      & $0.23$       & $0.29$      & $0.26$       & $0.19$  \\
            & $5\%$  & $0.56$      & $0.43$      & $0.29$       & $0.34$      & $0.34$       & $0.20$  \\
            & $10\%$ & $0.72$      & $0.54$      & $0.34$       & $0.41$      & $0.49$       & $0.25$  \\
      \hline
      $3$   & $1\%$  & $0.29$      & $0.28$      & $0.23$       & $0.28$      & $0.21$       & $0.19$  \\
            & $2\%$  & $0.36$      & $0.31$      & $0.28$       & $0.29$      & $0.26$       & $0.19$  \\
            & $5\%$  & $0.50$      & $0.40$      & $0.34$       & $0.36$      & $0.34$       & $0.23$  \\
            & $10\%$ & $0.66$      & $0.51$      & $0.41$       & $0.43$      & $0.49$       & $0.29$  \\
    \end{tabular}
  \end{center}
\end{table}

\begin{table*}
  \begin{center}
    \caption{Wave number $k$ [$h/$Mpc] below which each perturbation theory
      is accurate to the given accuracy level (Acc.) for the models in 
      Table~\ref{t:models}.  ``NL Kaiser'' uses the Time-RG
      real-space power spectrum in the Kaiser RSD 
      formula, Eq.~(\ref{e:Kaiser_lin}).
      \label{t:PTerrorOther}} 
    \tabcolsep=0.2cm
    \begin{tabular}{lc||cc|cc||cc|cc||cc|cc}
            &        &\multicolumn{4}{c}{\MooiNo}&\multicolumn{4}{c}{\MoooNi}&\multicolumn{4}{c}{\MoiiNi}\\
      $z$   & Acc.   &\multicolumn{2}{c}{Time-RG}&\multicolumn{2}{c}{NL Kaiser}&\multicolumn{2}{c}{Time-RG}&\multicolumn{2}{c}{NL Kaiser}&\multicolumn{2}{c}{Time-RG}&\multicolumn{2}{c}{NL Kaiser}\\
            &        & $\ell=0$    & $\ell=2$    & $\ell=0$     & $\ell=2$    & $\ell=0$     & $\ell=2$ & $\ell=0$    & $\ell=2$    & $\ell=0$     & $\ell=2$    & $\ell=0$     & $\ell=2$\\
      \hline
      \hline
      0  &  1\% & 0.11 & 0.06 & 0.04 & 0.10 & 0.11 & 0.06 & 0.05 & 0.12 & 0.12 & 0.06 & 0.06 & 0.11 \\ 
      &  2\% & 0.12 & 0.07 & 0.04 & 0.10 & 0.13 & 0.06 & 0.06 & 0.12 & 0.13 & 0.06 & 0.06 & 0.11 \\ 
      &  5\% & 0.16 & 0.07 & 0.12 & 0.10 & 0.16 & 0.06 & 0.16 & 0.13 & 0.15 & 0.08 & 0.15 & 0.11 \\ 
      & 10\% & 0.20 & 0.07 & 0.21 & 0.11 & 0.25 & 0.11 & 0.27 & 0.14 & 0.25 & 0.11 & 0.26 & 0.13 \\ 
      \hline
      0.5  &  1\% & 0.16 & 0.10 & 0.07 & 0.15 & 0.13 & 0.06 & 0.05 & 0.06 & 0.14 & 0.08 & 0.04 & 0.04 \\ 
      &  2\% & 0.17 & 0.10 & 0.07 & 0.15 & 0.15 & 0.11 & 0.05 & 0.06 & 0.15 & 0.08 & 0.04 & 0.06 \\ 
      &  5\% & 0.22 & 0.10 & 0.22 & 0.16 & 0.21 & 0.12 & 0.32 & 0.14 & 0.21 & 0.11 & 0.35 & 0.06 \\ 
      & 10\% & 0.25 & 0.15 & 0.33 & 0.16 & 0.34 & 0.49 & 1.17 & 0.17 & 0.33 & 0.39 & 1.12 & 0.16 \\ 
      \hline
      1  &  1\% & 0.18 & 0.10 & 0.07 & 0.09 & 0.19 & 0.17 & 0.06 & 0.06 & 0.15 & 0.13 & 0.06 & 0.06 \\ 
      &  2\% & 0.18 & 0.10 & 0.07 & 0.09 & 0.19 & 0.17 & 0.06 & 0.06 & 0.21 & 0.17 & 0.06 & 0.06 \\ 
      &  5\% & 0.25 & 0.16 & 0.47 & 0.18 & 0.36 & 0.45 & 0.70 & 0.15 & 0.89 & 0.46 & 0.52 & 0.16 \\ 
      & 10\% & 0.38 & 0.39 & 1.20 & 0.19 & 1.07 & 0.52 & 0.92 & 0.20 & 1.00 & 0.52 & 0.75 & 0.19 \\ 
      \hline
      2  &  1\% & 0.25 & 0.41 & 0.12 & 0.09 & 0.45 & 0.30 & 0.06 & 0.08 & 0.37 & 0.25 & 0.12 & 0.08 \\ 
      &  2\% & 0.25 & 0.55 & 0.36 & 0.18 & 0.49 & 0.35 & 0.26 & 0.08 & 0.44 & 0.31 & 0.27 & 0.13 \\ 
      &  5\% & 1.06 & 0.83 & 0.53 & 0.19 & 0.62 & 0.47 & 0.36 & 0.20 & 0.54 & 0.37 & 0.35 & 0.21 \\ 
      & 10\% & 1.13 & 1.10 & 0.85 & 0.25 & 0.79 & 0.62 & 0.53 & 0.26 & 0.69 & 0.54 & 0.49 & 0.26 \\ 
      \hline
      3  &  1\% & 0.62 & 0.39 & 0.26 & 0.19 & 0.36 & 0.34 & 0.26 & 0.20 & 0.37 & 0.37 & 0.23 & 0.13 \\ 
      &  2\% & 0.69 & 0.44 & 0.33 & 0.19 & 0.45 & 0.38 & 0.26 & 0.20 & 0.44 & 0.37 & 0.28 & 0.21 \\ 
      &  5\% & 0.86 & 0.58 & 0.44 & 0.25 & 0.60 & 0.51 & 0.38 & 0.25 & 0.58 & 0.52 & 0.40 & 0.26 \\ 
      & 10\% & 1.05 & 0.75 & 0.65 & 0.31 & 0.77 & 0.65 & 0.55 & 0.32 & 0.74 & 0.65 & 0.57 & 0.33 \\ 
    \end{tabular}
  \end{center}
\end{table*}

Table~\ref{t:PTerrorM000n0} uses the data from Fig.~\ref{f:multipoles_M000n0} to list the maximum $k$ up to which each perturbative calculation agrees with simulations to a given accuracy level.  The results follow the broad patterns expected from Fig.~\ref{f:P00_M000n0}: for percent-level accuracy at $z \lesssim 0.5$, the closure-theory-based approach of TNS is somewhat better than Time-RG, while for $5\%$-$10\%$ accuracy levels and at higher $z$, Time-RG is preferred.  In particular, we note that Time-RG diverges from the N-body power spectra rather smoothly, with no model-dependent catastrophic errors which could bias data analyses or forecasts.  Meanwhile, the non-linear Kaiser model, which uses the Time-RG non-linear power spectrum instead of the linear one in Eq.~(\ref{e:Kaiser_lin}), fails to match accurately the N-body quadrupole even at $z \sim 1$.  

\begin{figure}[tp]
  \hskip-3.2in{\Large{(a)}}
  
  \vskip-0.25in
  \includegraphics[width=3.4in]{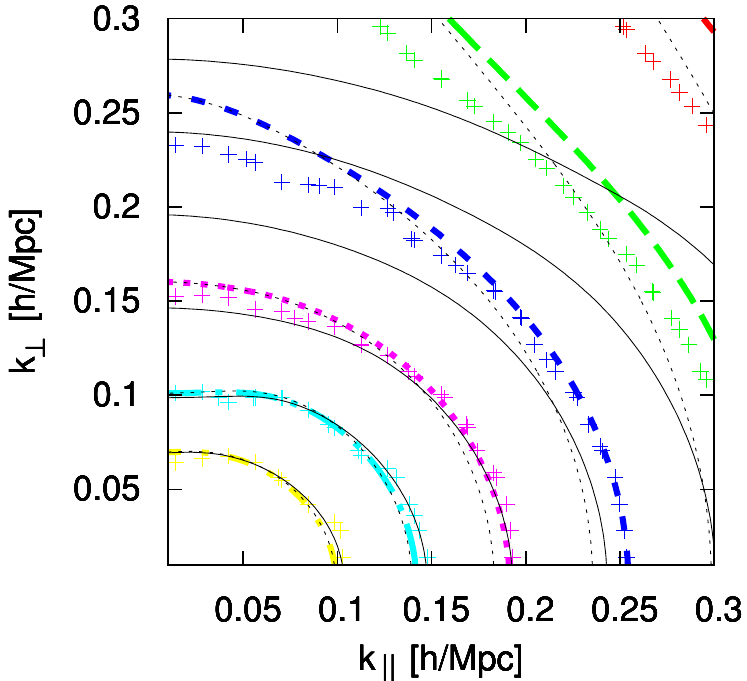}%
  
  \hskip-3.2in{\Large{(b)}}
  
  \vskip-0.25in
  \includegraphics[width=3.4in]{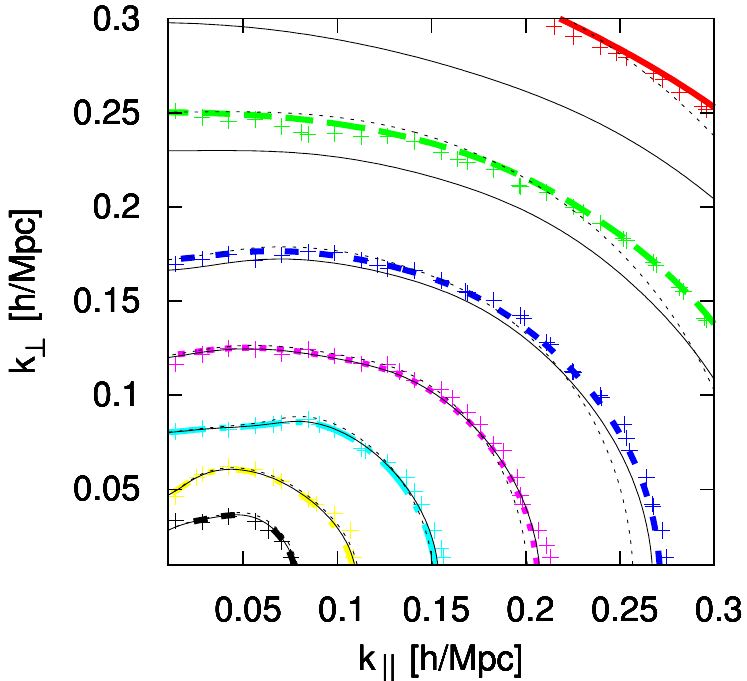}%
  \caption{
    Logarithmic contours of the redshift-space power spectrum
    $P_s(k_\parallel,k_\perp)$ for the massless-neutrino 
    model~\MoooNo, at (a)~$z=0$; (b)~$z=1$.  Points show N-body calculations, 
    thick dashed (colored) lines show Time-RG, solid black lines show TNS,
    and thin dashed black lines show the non-linear Kaiser approximation.
    \label{f:Ps_Nbody_M000n0}
  }
\end{figure}

Similar patterns can be seen in Figure~\ref{f:Ps_Nbody_M000n0}, which shows the two-dimensional power spectrum $P_s(k_\parallel,k_\perp)$ for simulations, Time-RG, TNS, and the non-linear Kaiser model.  The trends seen in Fig.~\ref{f:multipoles_M000n0} are evident here: Time-RG overestimates the monopole, particularly at $z=0$; TNS underestimates the monopole; and the Kaiser model substantially underestimates the quadrupole, even at $z=1$.  

\subsection{Evolving dark energy and massive neutrinos} 
\label{subsec:evolving_dark_energy_and_massive_neutrinos}

\begin{figure*}[tpb]
  \hskip-6.85in{\Large{(a)}}

  \vskip-0.25in
  \includegraphics[width=3.5in]{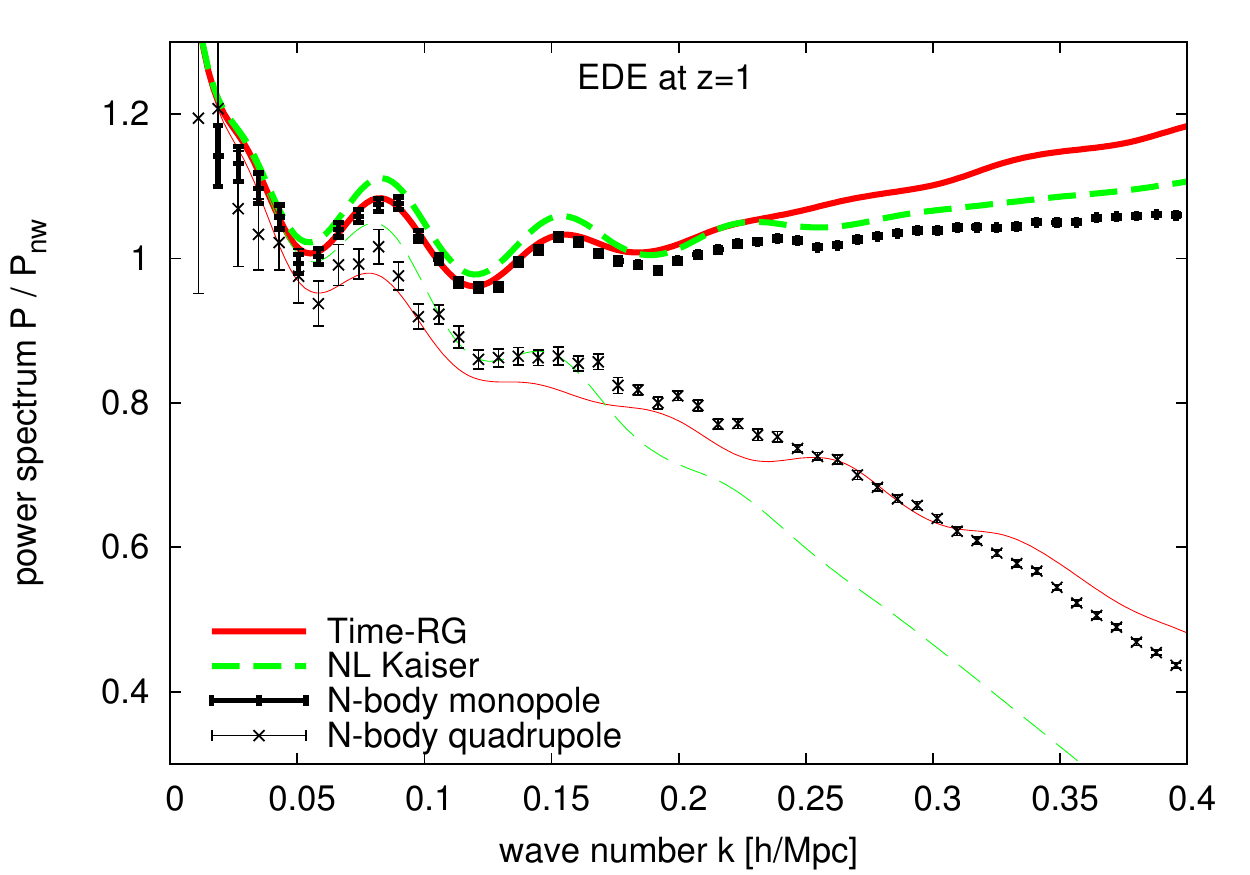}
  \includegraphics[width=3.5in]{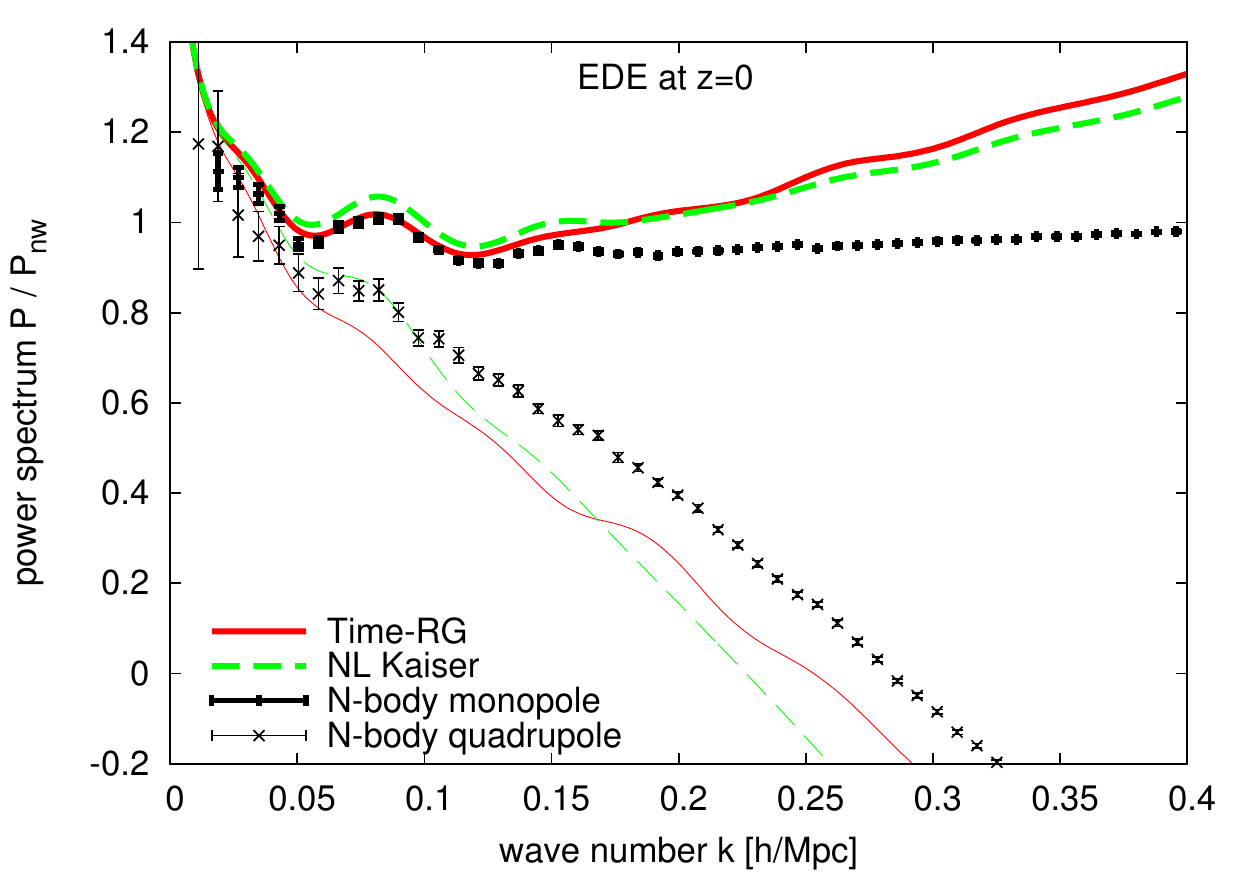}
  \vskip0.25in

  \hskip-6.85in{\Large{(b)}}

  \vskip-0.25in
  \includegraphics[width=3.5in]{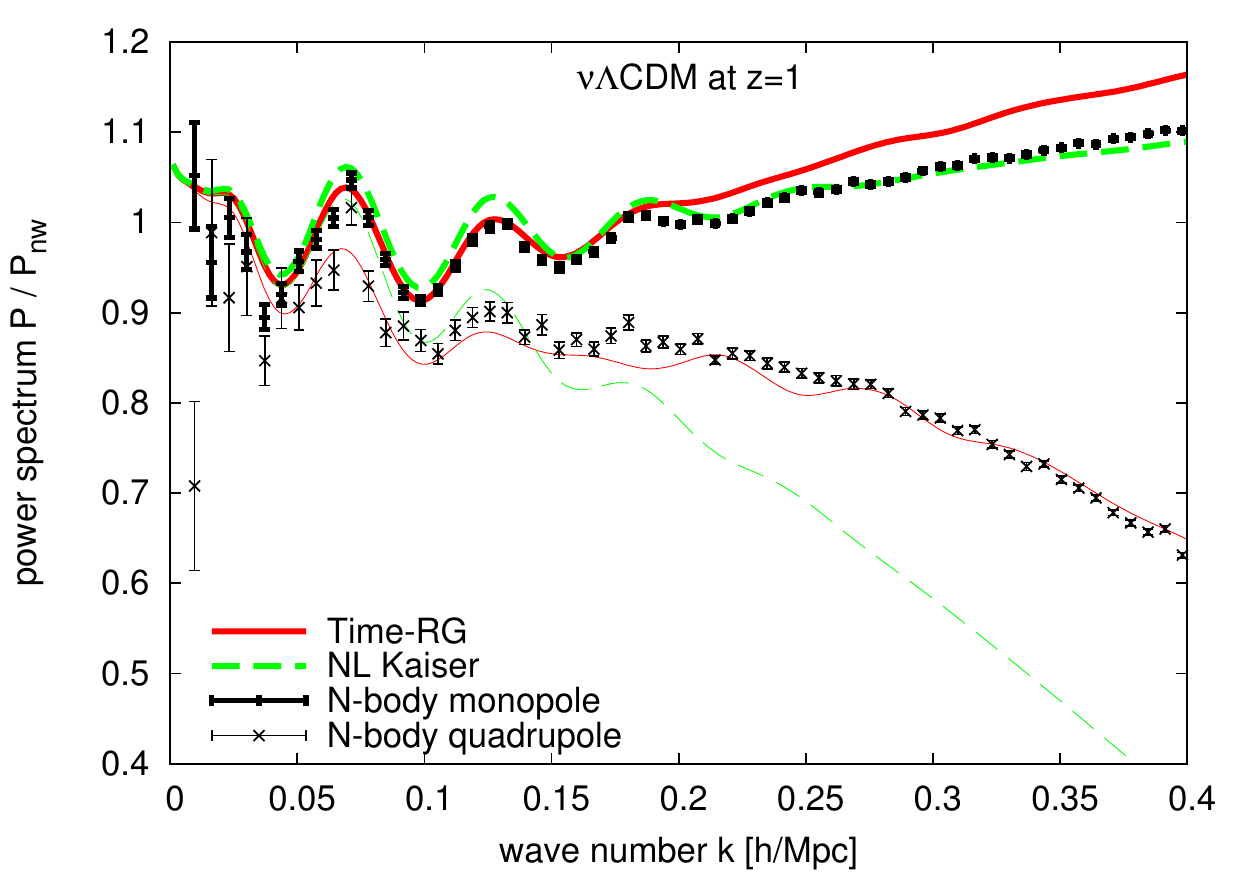}
  \includegraphics[width=3.5in]{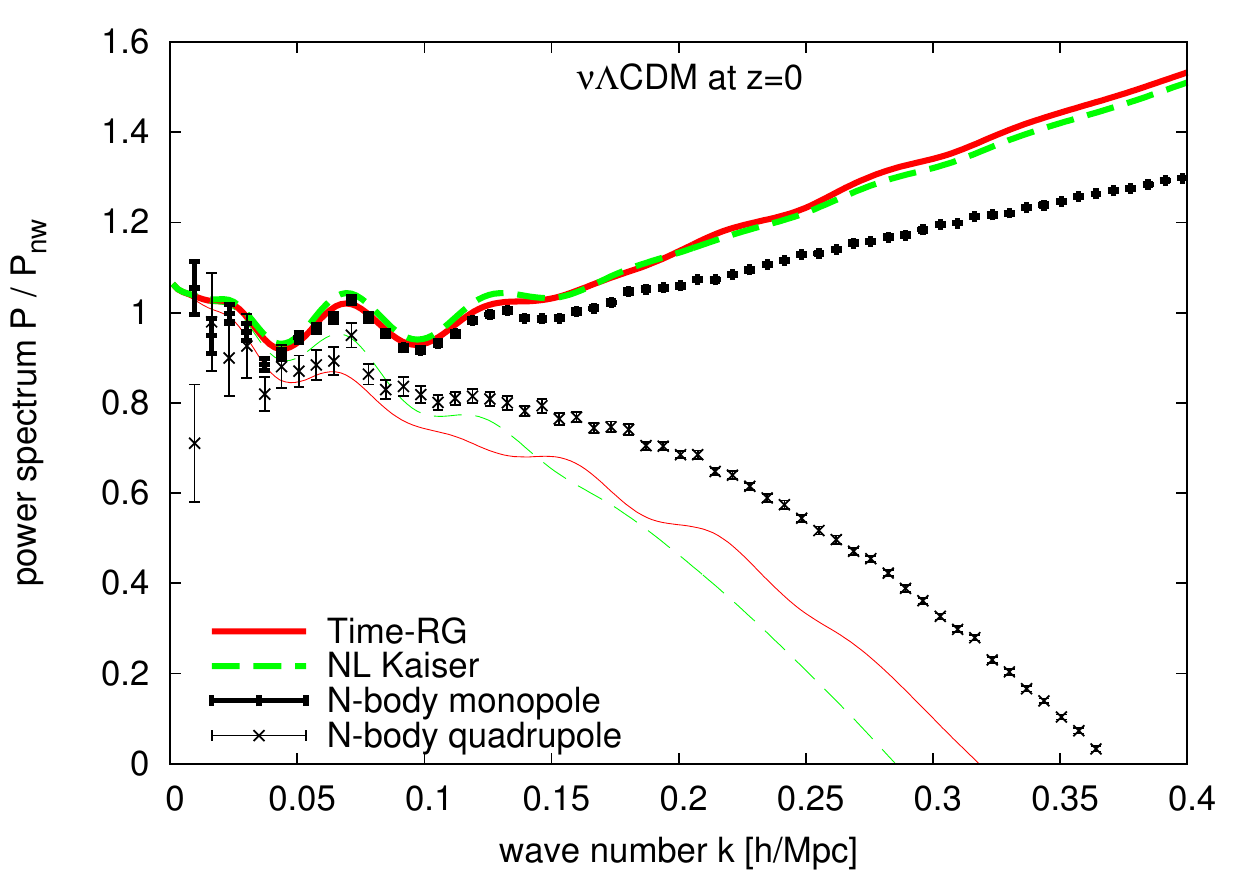}
  \vskip0.25in

  \hskip-6.85in{\Large{(c)}}

  \vskip-0.25in
  \includegraphics[width=3.5in]{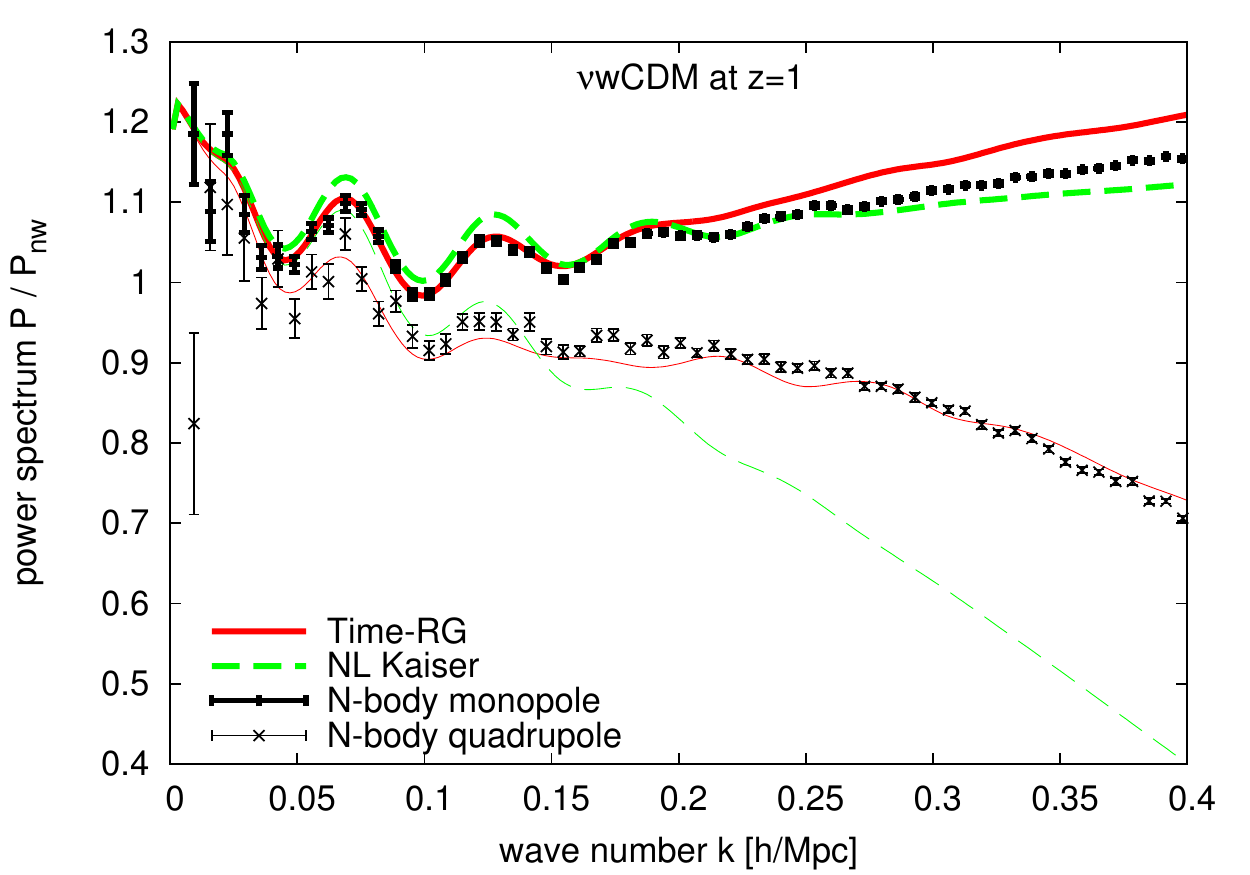}
  \includegraphics[width=3.5in]{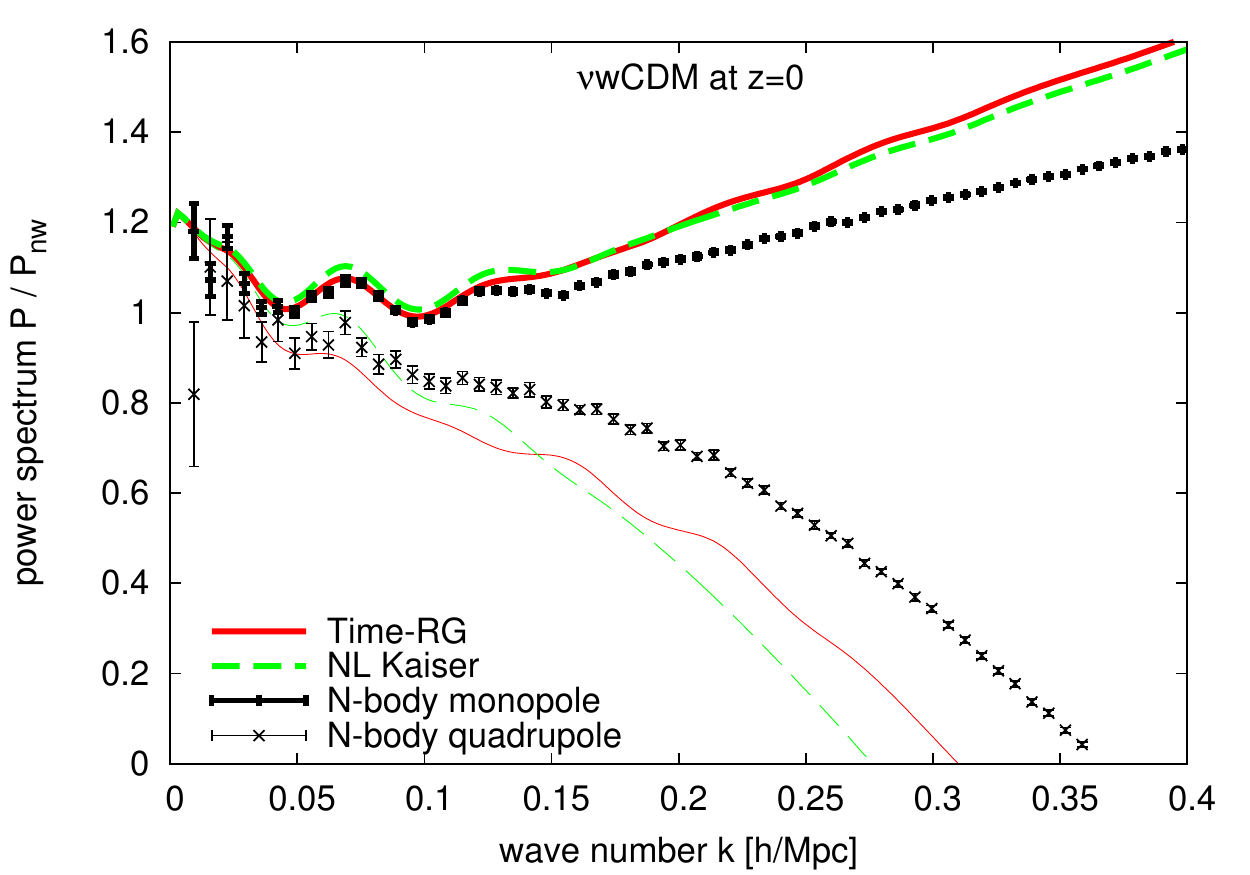}

  \caption{
    Monopoles (thick lines) and quadrupoles (thin lines) 
    of the redshift-space power spectra 
    at (left)~$z=1$ and (right)~$z=0$ for
    (a)~\MooiNo, early dark energy with massless neutrinos;
    (b)~\MoooNi, with $\sum m_\nu = 0.094$~eV;
    (c)~\MoiiNi, a rapidly-evolving dark energy with $\sum m_\nu = 0.29$~eV. 
    In each case, points represent N-body calculations, solid (red) lines
    show the Time-RG calculation, and dashed (green) lines show the
    non-linear Kaiser model.
    \label{f:rsd_with_de_and_mnu}
  }
\end{figure*}

Finally, we apply our calculation to the last three models in Table~\ref{t:models}: an early dark energy, a massive-neutrino $\Lambda$CDM model, and a dark energy with $w_a=-1.11$ and massive neutrinos.  Since next-generation galaxy surveys will go beyond the BOSS redshift~\cite{Anderson_2014} of $z=0.57$ to $z \approx 1$, we focus here on $z=1$ and $0$. Evidently from Figure~\ref{f:rsd_with_de_and_mnu}, the Time-RG RSD calculation presented here agrees well with N-body calculations of the monopole, quadrupole, and two-dimensional redshift-space power spectrum for a wide range of models at $z=1$.  By contrast, the Kaiser RSD model, Eq.~(\ref{e:Kaiser_lin}) applied to the non-linear real-space power spectrum, substantially underpredicts the quadrupole at quasi-linear scales.  At $z=0$, neither perturbative calculation is accurate beyond $k \approx 0.1~h/$Mpc for any of the models.  Table~\ref{t:PTerrorOther} lists the accuracy of both perturbative methods for all three models over the redshift range $0 \leq z \leq 3$.

\section{Conclusions}
\label{sec:conclusions}

Cosmological surveys over the next decade will measure the sum of neutrino masses, and should either detect or decisively exclude order-unity variations in the dark energy equation of state.  Observations are made in redshift space, with the redshift dependent on the line-of-sight velocity of an object as well as its distance.  Since peculiar velocities are sourced by overdensities, redshift-space distortions provide additional information about the scale-dependent growth of large-scale structure.  Perturbative techniques for calculating the redshift-space power spectrum in models with scale-independent growth were reviewed in Ref.~\cite{Kwan_2012}, with the method by Taruya, Nishimichi, and Saito~\TNS~(TNS) proving the most effective.

In this article we have extended the TNS approach to models with scale-dependent growth using the Time-RG perturbation theory, in which higher-order contributions to the power spectrum are described in terms of the bispectrum.  We have decomposed the higher-order corrections found by TNS into integrals $\Qlabc(k)$ over the bispectrum.  Using the Time-RG framework, we have derived the evolution equations for the $\Qlabc$ and computed the redshift-space power spectrum in models with massive neutrinos as well as rapidly-evolving dark energy.  Finally, we have confirmed the accuracy of our calculations by comparing them to N-body simulations conducted using the HACC code.  We compare our results to TNS and simulations in Figures~\ref{f:multipoles_M000n0} and \ref{f:Ps_Nbody_M000n0} as well as in Table~\ref{t:PTerrorM000n0}.  Figure~\ref{f:rsd_with_de_and_mnu} and Table~\ref{t:PTerrorOther} show that our results agree closely with N-body simulations for a wide range of models with massive neutrinos and rapidly-evolving dark energy equations of state.

Our work is applicable in a variety of ways.  Since our results are accurate at the $10\%$ level over a fairly large range of wave numbers, particularly for $z \gtrsim 1$, they can be used to forecast constraints from large-scale structure surveys.  Forecasts based upon linear perturbation theory are truncated at $k \sim 0.1~h/$Mpc, since linear theory at smaller scales overestimates the amount of information available from the BAO peak~\cite{Font-Ribera_2014}.  Accuracy can be improved by building an emulator combining our perturbative calculation at large scales with interpolated N-body power spectra at small scales~\cite{Heitmann_2014,Kwan_2013}.  Finally, perturbative techniques allowing for scale-dependent growth are applicable to modified gravity as well as to massive neutrinos.

\appendix
\section{Evolution of $\Pbis$ in Time-RG}
\label{sec:evolution_of_Pbis}

Rather than computing the full functional dependence of the bispectrum $B(k,q,p)$ on three different wave numbers, we may decompose $\Pbis(k,\mu)$ into a set of $k$-dependent bispectrum integrals whose time-evolution may be computed within the Time-RG framework.  Expanding the expression~(\ref{e:Pbis_tns}), we have
\begin{eqnarray}
\frac{\Pbis}{k\mu}
&=&
-\int \frac{d^3q}{(2\pi)^3} \frac{\muv{q}}{q}
\Big[
  \Bdtd(k,q,p_+) 
\nonumber\\
&&
  - \mu^2 \Bttd(k,q,p_+)
  - \muv{p_+}^2 \Bdtt(k,q,p_+)
\nonumber\\
&&
  + \mu^2 \muv{p_+}^2 \Bttt(k,q,p_+)
\Big]
- \plustominus
\end{eqnarray}
where $\vec p_\pm = \vec k \pm \vec q$ implies $p_\pm^2 \muv{p_\pm}^2 = k^2\mu^2 + q^2 \muv{q}^2 \pm 2 k q \mu \muv{q}$.  The three-dimensional integral over $\vec q$ can be evaluated as the integral over the magnitude $q$, the angle $\alpha$ between $\vec q$ and $\vec k$, and the angle $\beta$ of $\vec q - (\hat q \cdot \hat k)\vec k$ in the plane perpendicular to $\vec k$.  Since $B_{abc}(\vec k_1,\vec k_2, \vec k_3)$ is invariant under rotation in $\vec k_i$-space, the only $\beta$-dependent quantities in the integral are factors of $\muv{q}$ raised to integer powers; thus, the integral over $\beta$ can be evaluated trivially.  

Define the quantities
\begin{eqnarray}
\Qlabc(k)
&=&
\int \frac{q^2 dq \sin(\alpha)d\alpha}{(2\pi)^3} 
\frac{k}{p_+^2}
\left(\frac{q}{k}\right)^{\sigma_\ell}
\Pleg_{\left|\ell\right|}(\cos\alpha) 
\nonumber\\
&&
\times
B_{abc}(k,q,p_+)
+ (-1)^\ell \plustominus
\label{e:Qlabc}
\\
\sigma_\ell
&=&
\begin{cases}
  \mathrm{sign}(\ell) & \text{ if $\ell$ is odd} \\
  0                   & \text{ if $\ell$ is even}
\end{cases}
\end{eqnarray}
for $\ell = -1$, $0$, $1$, $2$, and $3$, where the $\Pleg$ are Legendre polynomials.  Then 
\begin{eqnarray}
\frac{\Pbisj{2}}{\pi k}
&=&
-2\QQ{-1}{\delta\theta\delta}
- 2\QQ{1}{\delta\theta\delta}
- \frac{8}{3} \QQ{2}{\delta\theta\delta}
- \frac{4}{3} \QQ{0}{\delta\theta\delta}
\nonumber\\&&
+ \frac{4}{3} \QQ{2}{\delta\theta\theta}
- \frac{4}{3} \QQ{0}{\delta\theta\theta}
+ \frac{6}{5} \QQ{3}{\delta\theta\theta}
- \frac{6}{5} \QQ{1}{\delta\theta\theta}
\label{e:Pbis2}
\\
\frac{\Pbisj{4}}{\pi k}
&=&
-2 \QQ{-1}{\ttd}
- 2 \QQ{1}{\ttd}
- \frac{8}{3} \QQ{2}{\ttd}
- \frac{4}{3} \QQ{0}{\ttd}
\nonumber\\&&
- 2 \QQ{-1}{\dtt}
- 4 \QQ{2}{\dtt}
- 2 \QQ{3}{\dtt}
+ \frac{4}{3} \QQ{2}{\ttt}
\nonumber\\&&
- \frac{4}{3} \QQ{0}{\ttt}
+ \frac{6}{5} \QQ{3}{\ttt}
- \frac{6}{5} \QQ{1}{\ttt}
\label{e:Pbis4}
\\
\frac{\Pbisj{6}}{\pi k}
&=&
-2 \QQ{-1}{\ttt}
- 4 \QQ{2}{\ttt}
- 2 \QQ{3}{\ttt}
\end{eqnarray}
where the $\Pbisj{j}$ and $\Qlabc$ are understood to depend on $k$.

Next, we study the evolution of the $\Qlabc$ in Time-RG.  Since the evolution of the $B_{abc}(k,q,p)$ only depends on the other components of $B_{abc}(k,q,p)$, the evolution of the $\Qlabc$ for fixed $\ell$ will only depend on the other components of $\Qlabc$.  We can easily multiply the bispectrum evolution equation~(\ref{e:eom_full_B}) by $(k/p_\pm^2) (q/k)^{\sigma_\ell} \Pleg_{|\ell|}(\cos\alpha)$, integrate over $d^3q$, and pull the time derivative outside of the integral.  Therefore,
\begin{eqnarray}
\partial_\eta \Qlabc
&=&
\!
-\Omega_{ad} \QQ{\ell}{dbc} \!
- \Omega_{bd} \QQ{\ell}{adc} \!
- \Omega_{cd} \QQ{\ell}{abd}
+ 2e^\eta \Rlabc \qquad\,
\label{e:eom_Q}
\\
\Rlabc\!(k)
&=&
\!
\int \!\frac{q^2 dq \sin\alpha d\alpha}{(2\pi)^3}
\frac{k}{p_+^2}
\left(\frac{q}{k}\right)^{\sigma_\ell}
\Pleg_{|\ell|}(\cos\alpha)
\nonumber\\&&
\!\times \Big[ 
  \gamma_{ade}(k,q,p_+) P_{db}(q) P_{ec}(p_+)
\nonumber\\&&
\!+ \gamma_{bde}(q,p_+,k)P_{dc}(p_+) P_{ea}(k)
\label{e:Rlabc}
\\&&
\!+ \gamma_{cde}(p_+,k,q) P_{da}(k) P_{eb}(q)
\Big]
+ (-1)^\ell \plustominus
\nonumber
\end{eqnarray}
where the $\Qlabc$ and $\Rlabc$ at each $\eta$ depend only on $k$. 

\section{Computation of $\Ptri$}
\label{sec:computation_of_Ptri}

Beginning with Eq.~(\ref{e:Ptri_tns}), we find $\Ptri$ by integrating over the angle $\beta$ of $\vec q$ in the plane perpendicular to $\vec k$.  Once again defining $\alpha$ as the angle between $\vec k$ and $\vec q$, and defining $\tau_{ab} = P_{a,\theta}(q) P_{b,\theta}(p_-)$, 
\begin{eqnarray}
\frac{\Ptrij{j}}{\pi k^2}
&=&
\int \frac{q^2 dq s_\alpha d\alpha}{(2\pi)^3 p_-^2}
\sum_{ab} S_{ab,j} \tau_{ab}
\quad \text{for }j=2,4,6,8 \qquad
\label{e:Ptrij}
\end{eqnarray}
where the $S_{ab,j}$ are given by:
\begin{eqnarray}
S_{\delta\delta,2}
&=&
-s_\alpha^2
\\
S_{\delta\theta,2}
&=&
-3 s_\alpha^4 q^2 / (4p_-^2)
\\
S_{\theta\delta,2}
&=&
-3 s_\alpha^6/4
\\
S_{\theta\theta,2}
&=&
-5 s_\alpha^6/8
\end{eqnarray}
\begin{eqnarray}
S_{\delta\delta,4}
&=&
2c_\alpha k/q + s_\alpha^2 - 2c_\alpha^2
\\
S_{\delta\theta,4}
&=&
-3s_\alpha^2 \tfrac{k^2}{p_-^2} + 9c_\alpha s_\alpha^2 \tfrac{kq}{p_-^2} 
+ s_\alpha^2(s_\alpha^2-4c_\alpha^2)\tfrac{3q^2}{2p_-^2}
\\
S_{\theta\delta,4}
&=&
3c_\alpha s_\alpha^2 k/q + 3s_\alpha^2(s_\alpha^2-4c_\alpha^2)/2
\\
S_{\theta\theta,4}
&=&
-s_\alpha^4 \tfrac{9k^2}{4p_-^2} + c_\alpha^2 s_\alpha^4 \tfrac{45kq}{4p_-^2}
+ s_\alpha^4(s_\alpha^2 - 6c_\alpha^2)\tfrac{15q^2}{8p_-^2}~~
\end{eqnarray}
\begin{eqnarray}
S_{\delta\theta,6}
&=&
c_\alpha^3/qp_-^2 + 3(s_\alpha^2-2c_\alpha^2)kq/p_-^2
\nonumber\\&&
+ 6(c_\alpha^2 s_\alpha^2 - c_\alpha^4/3 - s_\alpha^4/8) 6q^2/p_-^2
\\
S_{\theta\delta,6}
&=&
(2c_\alpha^2-3s_\alpha^2)k/q + 6(c_\alpha^2s_\alpha^2 - c_\alpha^4/3 - s_\alpha^4/8)
\qquad
\\
S_{\theta\theta,6}
&=&
3c_\alpha s_\alpha^2 k^3/(qp_-^2) + 9s_\alpha^2(s_\alpha^2-4c_\alpha^2)k^2/(4p_-^2)
\nonumber\\&&
+ 30(c_\alpha^3 s_\alpha^2 - c_\alpha s_\alpha^4/4)kq/p_-^2
\nonumber\\&&
+ 45s_\alpha^2(c_\alpha^2 s_\alpha^2 - 2c_\alpha^4/3 -s_\alpha^4/12)q^2/(2p_-^2)
\end{eqnarray}
\begin{eqnarray}
S_{\theta\theta,8}
&=&
(2c_\alpha^3 - 3c_\alpha s_\alpha^2)\tfrac{k^3}{qp_-^2}
+ 18(c_\alpha^2s_\alpha^2 - \tfrac{1}{3}c_\alpha^4 - \tfrac{1}{8}s_\alpha^4)
\tfrac{k^2}{p_-^2}
\nonumber\\&&
+ 6(c_\alpha^6 - 5c_\alpha^3 s_\alpha^2 + \tfrac{15}{8}c_\alpha s_\alpha^4)
\tfrac{kq}{p_-^2}
\nonumber\\&&
- (2c_\alpha^6 - 15c_\alpha^4 s_\alpha^2 
+ \tfrac{45}{4}c_\alpha^2 s_\alpha^4 - \tfrac{5}{8}s_\alpha^6)
\tfrac{q^2}{p_-^2}
\end{eqnarray}
We have used the shorthand notation $s_\alpha = \sin(\alpha)$ and $c_\alpha = \cos(\alpha)$.  All components of $S_{ab,j}$ not listed above are zero.

\section{Error bounds and $\sigma_v$ fits}
\label{sec:error_bounds_and_sigv_fits}

\subsection{Scale-dependence of $\OmegaTRG$} 

In the general multi-species case as well as in the linear neutrino approximation of Sec.~\ref{subsec:massive_neutrinos_linear_approximation}, the linear evolution matrix $\OmegaTRG$ is scale-dependent.  In particular, Eq.~(\ref{e:eom_Omega10}) shows that the fractional change in $\OmegaTRG_{10}$ is of order $\fnu$ in massive neutrino models.  This scale-dependence has been neglected in real-space and redshift-space Time-RG, Eqs.~(\ref{e:eom_I},\ref{e:eom_Q}), in which $\OmegaTRG$ has been pulled outside the integrals over wave number.  Here we estimate the error associated with this approximation and show that it is small enough to be negligible even for neutrino masses several times current bounds.

\begin{figure}[tp]
  \begin{center}
    \includegraphics[width=3.3in]{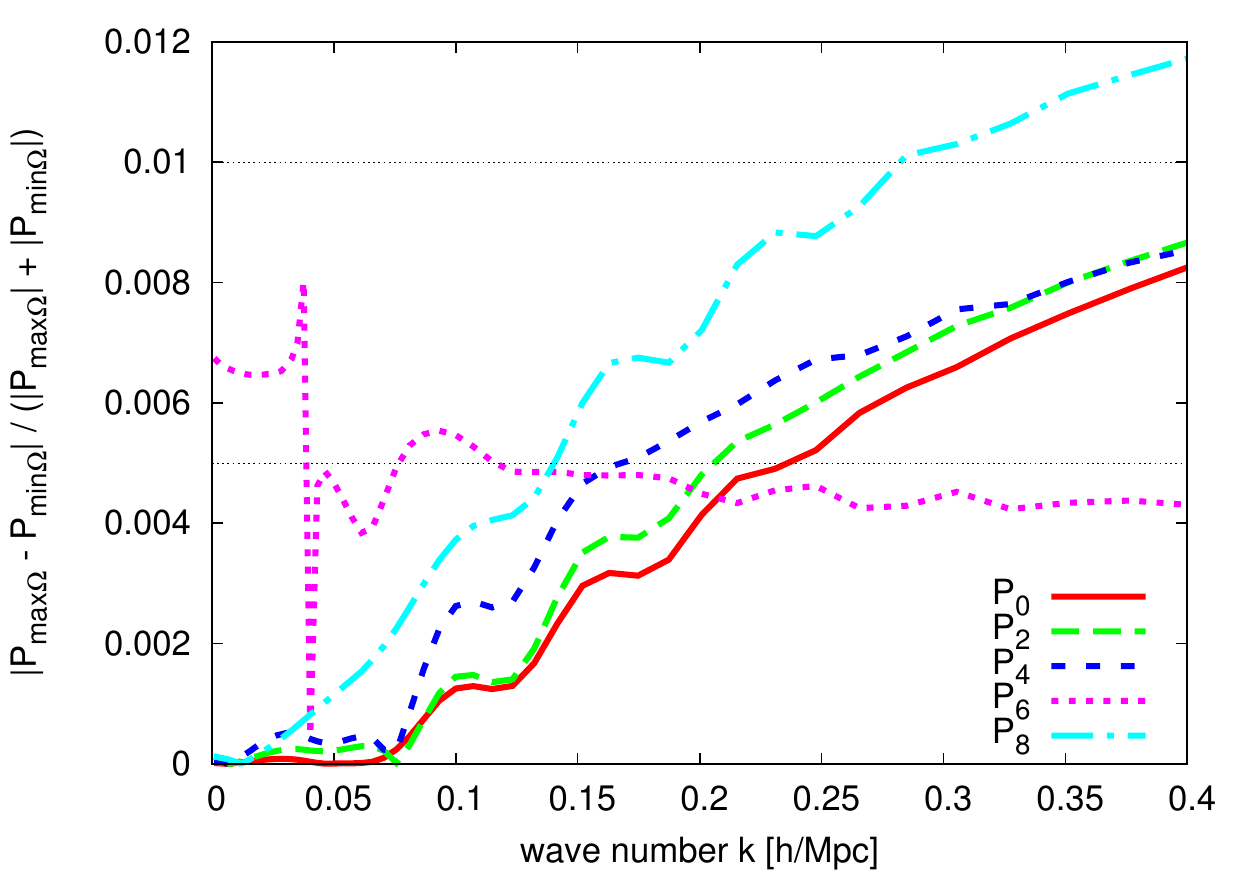}%
    \caption{
      Error associated with neglecting the $k$-dependence of $\OmegaTRG_{10}$, 
      Eq.~(\ref{e:eom_Omega10}), in the non-linear Time-RG corrections 
      Eqs.~(\ref{e:eom_I},\ref{e:eom_Q}).  Differences between power spectra
      computed with the maximum and minimum values of $\OmegaTRG_{10}$ are
      shown at $z=0$ for a model with $\omega_\nu=0.01$, corresponding
      to $\sum m_\nu = 0.94$~eV.
      \label{f:err_kdepOmega}
    }
  \end{center}
\end{figure}

\begin{figure}[htp]
  \begin{center}
    \includegraphics[width=3.3in]{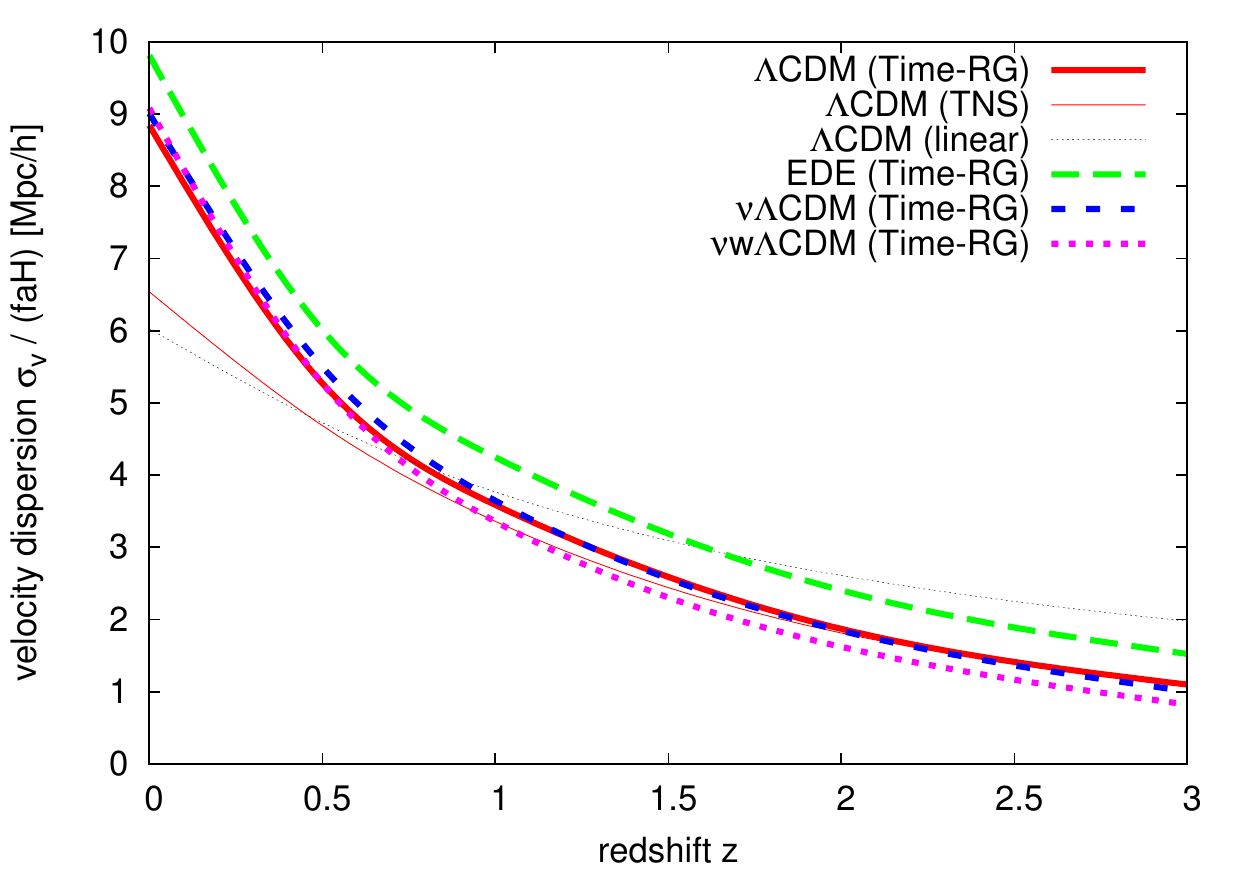}%
    \caption{
      Best-fit velocity dispersion $\sigma_v$ for the models in 
      Table~\ref{t:models}, fitting to a suite of $15$ Particle-Mesh
      N-body simulations at scales $k \leq k_{\max}=0.2~h/$Mpc.
      \label{f:sigv}
    }
  \end{center}
\end{figure}

In order to place an upper bound on this $k$-dependent $\OmegaTRG$ error, we study a model similar to~\MoooNi~but with ten times the neutrino content, $\omega_\nu = 0.01$, corresponding to $\sum m_\nu = 0.94$~eV.  The function $\OmegaTRG_{10}(z,k)$ drops from a maximum value at large scales, where neutrinos cluster like cold matter, to a small-scale value that is smaller by a fraction $\approx \fnu / (1-\fnu)$.  Moreover, we expect the error to be smaller at higher $z$, since it only affects non-linear correction terms, which grow with time.  

We estimate the $k$-dependent $\OmegaTRG$ error by computing power spectra using either the maximum or the minimum values of $\OmegaTRG_{10}$ in Eqs.~(\ref{e:eom_I},\ref{e:eom_Q}). Our estimate, half the difference between the two, is shown in Fig.~\ref{f:err_kdepOmega} for the power spectrum components $P_j(k)$ defined in Eq.~\ref{e:Pj}.  For the dominant contributors to the redshift-space power spectrum, $P_0$, $P_2$, and $P_4$, the error is less than $0.5\%$ for $k \leq 0.15~h/$Mpc and less than $0.9\%$ for $k \leq 0.4~h/$Mpc.  Errors in the smaller terms $P_6$ and $P_8$ are less than $1\%$ for $k \leq 0.25~h/$Mpc.  At higher $z$, and for more realistic $\fnu$, we expect these errors to be several times smaller, meaning that they are negligible compared with the errors listed in Tables~\ref{t:PTerrorM000n0} and \ref{t:PTerrorOther}.

\subsection{Best-fit $\sigma_v$}
\label{subsec:best-fit_sigmav}

Our results in Tables~\ref{t:PTerrorM000n0} and \ref{t:PTerrorOther} are based on fitting the velocity dispersion $\sigma_v$ in the Lorentzian streaming function $\Ffog(f\sigma_v k \mu)$ to the N-body simulations over the range $k \leq k_{\max}=0.2~h/$Mpc.  The resulting velocity dispersions are shown in Fig.~\ref{f:sigv} for the models in Table~\ref{t:models}.  Velocity dispersions for models~\MoooNo,~\MoooNi, and~\MoiiNi~are similar, while the early dark energy model~\MooiNo, which enhances structure growth, has a larger velocity dispersion.  Also shown, for~\MoooNo, are the velocity dispersion in the TNS calculation and that of linear theory, Eq.~(\ref{e:sigma_v}).  For~\MoooNo, both non-linear calculations find $\sigma_v$ below linear theory at high $z$ and above linear theory at low $z$.  The best-fit $\sigma_v$ values appear to be biased by perturbation theory errors at low $z$.  Time-RG, which overpredicts late-time small-scale power, has a higher $\sigma_v$ than TNS, which underpredicts it.

\subsection{Sensitivity to $k_{\max}$}
\label{subsec:sensitivity_to_kmax}

\begin{table}[tp]
  \begin{center}
    \caption{
      Sensitivity to $k_{\max}$.
      Wave numbers $k$ [$h/$Mpc] below which Time-RG perturbation theory is
      accurate to the given accuracy level (Acc.) for model~\MoooNi~are shown.  
      In each case, the velocity dispersion $\sigma_v$ in the streaming
      function $\Ffog(f\sigma_v k\mu)$ was fit to the N-body power spectrum
      over the range $k \leq k_{\max}$.  For comparison, 
      Tables~\ref{t:PTerrorM000n0} and \ref{t:PTerrorOther}
      used $k_{\max}=0.2~h/$Mpc.
      \label{t:vary_kmax}
    }
    \tabcolsep=0.2cm
    \begin{tabular}{lc|cc|cc|cc}
      $z$   & Acc.   &\multicolumn{2}{c}{$\frac{k_{\max}}{h/\mathrm{Mpc}}=0.1$}&\multicolumn{2}{c}{$\frac{k_{\max}}{h/\mathrm{Mpc}}=0.15$}&\multicolumn{2}{c}{$\frac{k_{\max}}{h/\mathrm{Mpc}}=0.25$}\\
            &        & $\ell=0$    & $\ell=2$                & $\ell=0$     & $\ell=2$               & $\ell=0$     & $\ell=2$\\
      \hline
      0  &  1\% & 0.09 & 0.06 & 0.11 & 0.06 & 0.13 & 0.06\\ 
      &  2\% & 0.11 & 0.06 & 0.13 & 0.06 & 0.13 & 0.06 \\ 
      &  5\% & 0.14 & 0.11 & 0.15 & 0.11 & 0.19 & 0.06 \\ 
      & 10\% & 0.21 & 0.17 & 0.21 & 0.11 & 0.27 & 0.11 \\ 
      \hline
      0.5  &  1\% & 0.13 & 0.11 & 0.13 & 0.11 & 0.14 & 0.06\\ 
      &  2\% & 0.13 & 0.12 & 0.13 & 0.11 & 0.19 & 0.06 \\ 
      &  5\% & 0.19 & 0.17 & 0.21 & 0.17 & 0.26 & 0.11 \\ 
      & 10\% & 0.30 & 0.36 & 0.32 & 0.40 & 0.42 & 0.17 \\ 
      \hline
      1  &  1\% & 0.16 & 0.17 & 0.17 & 0.17 & 0.19 & 0.14\\ 
      &  2\% & 0.19 & 0.17 & 0.19 & 0.17 & 0.21 & 0.17 \\ 
      &  5\% & 0.32 & 0.36 & 0.32 & 0.44 & 0.96 & 0.17 \\ 
      & 10\% & 1.09 & 0.45 & 1.07 & 0.49 & 1.04 & 0.59 \\ 
      \hline
      2  &  1\% & 0.19 & 0.19 & 0.41 & 0.17 & 0.41 & 0.29\\ 
      &  2\% & 0.28 & 0.21 & 0.49 & 0.30 & 0.49 & 0.30 \\ 
      &  5\% & 0.78 & 0.86 & 0.60 & 0.38 & 0.61 & 0.38 \\ 
      & 10\% & 0.96 & 0.99 & 0.77 & 0.58 & 0.77 & 0.59 \\ 
      \hline
      3  &  1\% & 0.19 & 0.18 & 0.36 & 0.34 & 0.35 & 0.34\\ 
      &  2\% & 0.28 & 0.19 & 0.45 & 0.38 & 0.45 & 0.35 \\ 
      &  5\% & 1.08 & 0.28 & 0.58 & 0.49 & 0.58 & 0.49 \\ 
      & 10\% & 1.14 & 0.53 & 0.76 & 0.64 & 0.76 & 0.62 \\ 
    \end{tabular}
  \end{center}
\end{table}

In the quasi-linear regime, errors in perturbation theory increase with increasing $k$, while uncertainties in N-body simulations and actual data decrease due to a greater number of modes.  Thus one might worry that our $\sigma_v$ fitting procedure is dominated by wave numbers near $k_{\max}$, which may be at the edge of the regime of validity of perturbation theory.  Table~\ref{t:vary_kmax} tests the sensitivity of our redshift-space Time-RG perturbative calculation to $k_{\max}$ for model~\MoooNi.  The table shows wave numbers up to which perturbation theory is accurate for several accuracy thresholds, for $k_{\max} = 0.1~h/$Mpc, $0.15~h/$Mpc, and $0.25~h/$Mpc.  Comparison to Table~\ref{t:PTerrorOther} shows that $k_{\max} = 0.15~h/$Mpc, $0.2~h/$Mpc, and $0.25~h/$Mpc are similar.  At the $1-2\%$ accuracy levels and at $z\geq 2$, $k_{\max} = 0.1~h/$Mpc leads to a significantly worse fit.  This is not surprising, since the N-body error bars in Fig.~\ref{f:rsd_with_de_and_mnu}~(b) are large for $k<0.1~h/$Mpc, and since not much non-linear information is available at such large scales.  We conclude that our results are stable over the range $0.15~h/\mathrm{Mpc} \leq k_{\max} \leq 0.25~h/$Mpc.

\subsection*{Acknowledgments}
We are grateful to D.~Chung, E.~Jennings, T.~Okumura, and M.~Takada for 
insightful discussions. 
The authors were supported by the U.S. Department of Energy, Basic Energy
Sciences, Office of Science, under contract No. DE-AC02-06CH11357.
This research used resources of the ALCF, which is supported by DOE/SC
under contract DE-AC02-06CH11357 and resources of the OLCF, which is
supported by DOE/SC under contract DE-AC05-00OR22725.

The submitted manuscript has been created by UChicago Argonne, LLC,
Operator of Argonne National Laboratory (``Argonne''). Argonne, a
U.S. Department of Energy Office of Science laboratory, is operated
under Contract No. DE-AC02- 06CH11357. The U.S. Government retains for
itself, and others acting on its behalf, a paid-up nonexclusive,
irrevocable worldwide license in said article to reproduce, prepare
derivative works, distribute copies to the public, and perform
publicly and display publicly, by or on behalf of the Government.

\bibliographystyle{unsrt}
\bibliography{time-rg_rsd}

\end{document}